\documentclass[12pt]{article}
\usepackage[final]{graphicx}
\usepackage{natbib}
\usepackage{amsfonts}
\usepackage{amsmath}
\usepackage{amssymb}
\usepackage[plain,noend]{algorithm2e}
\usepackage{multirow}
\usepackage{bm}
\usepackage{color}
\usepackage{bbm}
\usepackage{float} 
\usepackage{indentfirst}
\usepackage{setspace}
\usepackage{caption}
\usepackage{subcaption}
\usepackage[left=1cm, right=1cm, top=2cm, bottom=2cm]{geometry}
\UseRawInputEncoding

\newtheorem{definition}{Definition}[section]

\newtheorem{theorem}{Theorem}[section]

\newenvironment{example}[1][Example]{\begin{trivlist}
\item[\hskip \labelsep {\bfseries #1}]}{\end{trivlist}}

\newcommand{\bdelta}{\delta}  %
\newcommand{\GFD}{GFD\xspace}
\newcommand{\CDs}{CDs\xspace}
\newcommand{\CD}{CD\xspace}

\newcommand{\bX}{X}
\newcommand{\bT}{T}
\newcommand{\bY}{Y}
\newcommand{\by}{y}
\newcommand{\bU}{U}
\newcommand{\bu}{u}
\newcommand{\bG}{G}
\newcommand{\bv}{v}
\newcommand{\bV}{V}
\newcommand{\btheta}{\theta}
\newcommand{\bTheta}{\Theta}
\newcommand{\bz}{z}
\newcommand{\bZ}{Z}

\newcommand{\etab}{\eta}
\newcommand{\bd}{d}
\newcommand{\bi}{{\boldsymbol{i}}}
\newcommand{\argmin}{\operatornamewithlimits{arg\ min}}

\newcommand{\yifan}[1]{\textcolor{green}{#1}\xspace}

\begin{document}
\onehalfspacing
\title{Confidence Distribution and Distribution Estimation \\ for Modern Statistical Inference}
\author{Yifan Cui, Min-ge Xie\\National University of Singapore, Rutgers University}

\date{}
\maketitle

\abstract{

This paper introduces to readers the new concept and methodology of confidence distribution and the modern-day distributional inference in statistics. This discussion should be of interest to people who would like to go into the depth of the statistical inference methodology and to utilize distribution estimators in practice.
We also include in the discussion the topic of generalized fiducial inference, a special type of modern distributional inference, and relate it to the concept of confidence distribution. Several real data examples are also provided for practitioners. We hope that the selected content covers the greater part of the developments on this subject.

}

\vspace{0.8cm}

\textbf{\large Keywords}
\begin{itemize}
    \item \textcolor{blue}{Confidence distributions}
    \item \textcolor{blue}{Fiducial inference}
    \item \textcolor{blue}{Distributional inference}
    \item \textcolor{blue}{Confidence intervals}
    \item \textcolor{blue}{Coverage}
    \item \textcolor{blue}{Fusion learning}
    \item \textcolor{blue}{Bayesian, fiducial, and frequentist~(BFF)}
\end{itemize}

\tableofcontents

\section{Introduction}

A \yifan{confidence distribution} (\CD) refers to a sample-dependent distribution function that can represent confidence intervals (regions) of all levels for a parameter of interest \citep{XieSingh2013, schweder2016confidence}.
Instead of the usual point estimator or confidence interval,  \CD is a distribution estimator of a parameter of interest with a pure frequentist interpretation.
The development of the \CD can be traced back to, for example, \cite{Fisher1930,neymann1941,cox1958,lehmann1993}. However, its associated inference schemes and applications have not received much attention until the recent surge of interest in the research of \CD and its applications \citep{Efron1998,SchwederHjort2002,SchwederHjort2003,schweder2016confidence,XieSinghStrawderman2011,SinghXieStrawderman2005,singhxiestrawderman2007,XieSingh2013,lawless2005,tian2011,yang2016,liu2014exact,Liu2015MultivariateMO}.
All of these developments of \CDs, along with a modern definition and interpretation, provide a powerful inferential tool for statistical inference.

One of the main contributions of \CD is its applications on \yifan{fusion learning} \citep{liusingh1997,SchwederHjort2002,SinghXieStrawderman2005,tian2008exact,
XieSinghStrawderman2011, HannigXie2012, XieEtAl2013,liu2014exact,ChenXie2014,claggett2014,Liu2015MultivariateMO,shen2019jasa}. Combining \CDs from independent studies naturally preserves more information from the individual studies than a traditional approach of combining only point estimators. A unified framework of combining \CDs for fusion learning generally includes three steps: 1) using a CD to summarize relevant information or obtain an inference result from each study; 2) combining information
from different sources or studies by combining these \CDs;
3) making inference via the combined CD. This approach has sound theoretical support and has been applied to many practical situations with much success.

On a different note, the \yifan{fiducial distribution} may be considered as one special type of \CD, which provides a systematic way to obtain a
\CD. The origin of fiducial inference can be traced back to R.A. Fisher \citep{Fisher1930} who introduced the concept of
a fiducial distribution for one parameter, and proposed the use of this fiducial distribution to avoid the problems related to the choice of a prior distribution.
Since the mid 2000s, there has been a renewed interest in modifications of fiducial inference \citep{WangYH2000, SchwederHjort2003, Fraser2004, Fraser:2005tc,XuGuoying2006, Dempster2008,FraserNaderi2008, Hannig2009, EdlefsenLiuDempster2009,BergerBernardoSun2009,FraserFraserStaicu2010, FraserReidMarrasYi2010, MartinZhangLiu2010,ZhangLiu2011, Fraser2011,BayarriEtAl2012,BergerBernardoSun2012,TaraldsenLindquist2013,XieSingh2013,MartinLiu2013a,
MartinLiu2013b,
MartinLiu2013c, VeroneseMelilli2014, martin2015inferential,schweder2016confidence,hannig2016generalized,ryan2017IM,MARTIN2018105,qiu2018}.

We briefly overview these modern approaches which extend Fisher's original fiducial argument. We then focus on a recent development termed \yifan{generalized fiducial inference} and its applications \citep{Hannig2009,CisewskiHannig2012,WandlerHannig2012b,IVERSON2014GFI, lai2015,hannig2016generalized,liu2016generalized,liu2017generalized,williams2019,williams2019b,Cui2019,Neupert2019} that greatly expand the applicability of fiducial ideas.
We demonstrate this recipe on several examples of varying complexity. The statistical procedures derived by the generalized fiducial inference often have very good performance from both theoretical and numerical points of view.

\section{Confidence distribution}

\subsection{The concept of \CD \label{sec:cd}}

This section will mainly focus on the concept of \CD. The \CD can be viewed as a \yifan{distribution estimator}, which can be utilized for constructing statistical procedures such as point estimates, \yifan{confidence intervals}, hypothesis tests, etc.
The basic notion of \CDs is related to the fiducial distribution of \cite{Fisher1930}, however, it is a pure frequentist concept. Some have suggested to view \CD as the frequentist analogue of Bayesian posterior distribution \cite[e.g.][]{SchwederHjort2003, schweder2016confidence}. More broadly, if the credible intervals or regions obtained from a Bayesian posterior match with frequentist intervals or regions (either exactly or asymptotically), then the Bayesian posterior can be viewed as \CD  and thus Bayesian approach is also a way to obtain \CD \citep{XieSingh2013}.

Suppose $X_1, , X_2,\ldots,X_n$ are independent and identically distributed, and $\mathcal  X$ is the sample space corresponding to the dataset 
$(X_1, X_2,\ldots,X_n)$. Let $\theta$ be a scalar parameter of interest, and $\Theta$ be the parameter space.
The following formal definitions of \CD and asymptotic \CD are proposed in \cite{SchwederHjort2002,SinghXieStrawderman2005}.

\begin{definition}[CD and asymptotic CD]

A function $H_n(\cdot)=H_n(x,\cdot)$ on $\mathcal X \times \Theta \rightarrow [0,1]$ is called a \CD for a parameter, if (1) For each given $x \in \mathcal X$, $H_n(\cdot)$ is a (continuous) cumulative distribution function
on $\Theta$;  (2) At the true parameter value $\theta=\theta_0$, $H_n(\theta_0) \equiv H_n(x,\theta_0)$, as a function of the sample
$x$, follows the uniform distribution $U(0, 1)$.
In addition, the function $H_n(\cdot)$ is called an asymptotic \CD if condition (2) is replaced by (2') At the true parameter $\theta=\theta_0$, $H_n(\theta_0) \overset{d}{\to}  U(0,1)$ as $n \rightarrow \infty$.

\end{definition}

From a non-technical point of view, a \CD is a function of both the parameter and the sample which satisfies two conditions. The first condition basically states that for any fixed sample, a
\CD is a distribution function on the parameter space. The second condition essentially requires that the corresponding inference derived by a \CD has desired frequentist properties.
Section~\ref{sec:inference} will further discuss how to use the second condition to extract information from a \CD to make inference.

\cite{Birnbaum1961} introduced the concept of confidence curve as  ``an omnibus technique for estimation and testing statistical hypotheses,'' which was independent of the development of CD.
From a CD $H_n(\theta)$, the \yifan{confidence curve} can be written as $$CV_n(\theta)=2\min\{ H_n(\theta),1-H_n(\theta)\}.$$
Indeed, confidence curve is an alternative expression of \CD and it is a very useful graphical tool for visualizing CDs.
On a plot of $CV_n(\theta)$ versus $\theta$, a line across the $y$-axis of the significance level $\alpha$, for any $0 < \alpha < 1$, intersects with the confidence curve at two points, and these two points correspond to an $1-\alpha$ level, equal-tailed, two-sided confidence interval for $\theta$. In addition, the maximum of a confidence curve is the median of the \CD which is the recommended point estimator.

We present below five illustrating examples of \CDs.
More examples refer to \cite{SinghXieStrawderman2005,XieSingh2013,schweder2016confidence}.

\begin{figure}[!h]
\begin{center}
{\includegraphics[width=18cm]{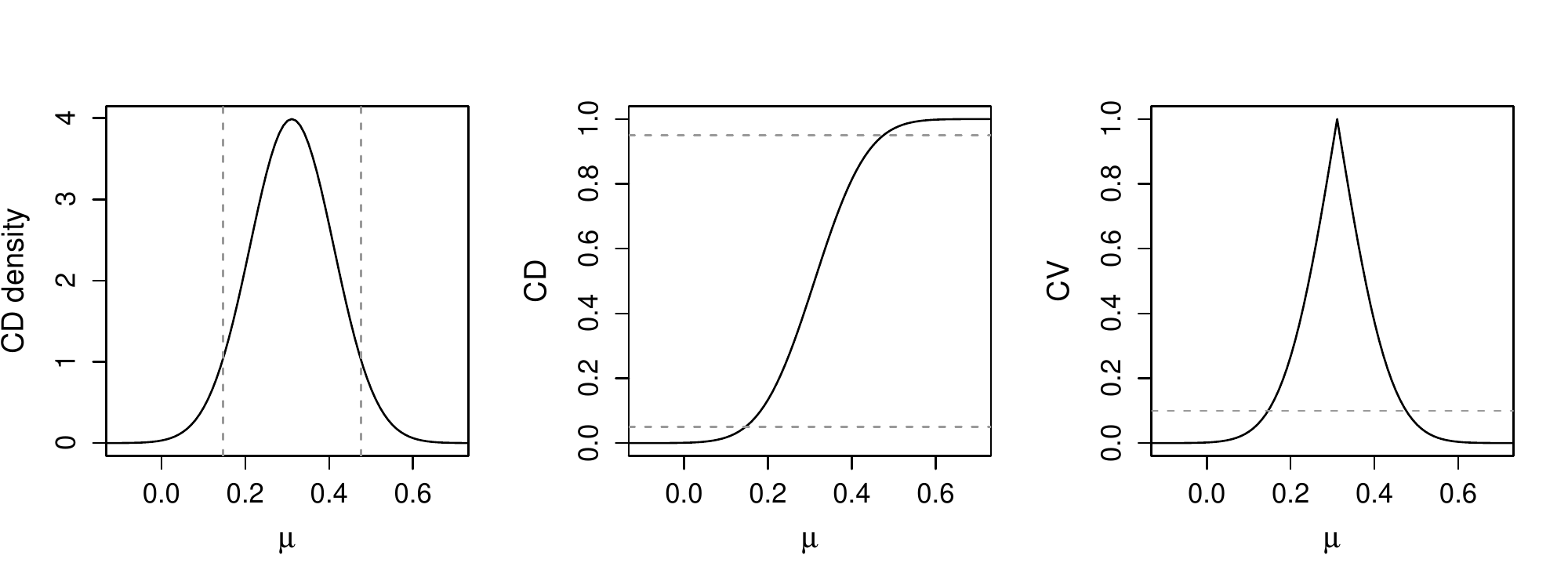}}
\end{center}
\caption{Confidence distribution presented in Example 1 in the forms of density function, cumulative distribution function, and confidence curve. }
\label{pic:cc}
\end{figure}

\begin{example}[Example 1]
Suppose the data $X_i \sim N(\mu,1), i = 1,\ldots, n$, with unknown $\mu$. Let $\bar x_n$ denote the sample mean. Then $N(\bar x_n, 1/n )$ is a \CD for $\mu$, and it can be represented in the following three forms:
(i) Confidence distribution (cumulate distribution form): $H_n(\mu) = \Phi(\sqrt n (\mu- \bar x_n) )$; (ii) \yifan{Confidence density} (density form): $h_n(\mu) = \frac{1}{\sqrt{2\pi/n}} \exp\{ -\frac{n}{2}(\mu- \bar x_n)^2 \}$;
(iii) Confidence curve: $CV_n(\mu)=2\min\{ \Phi(\sqrt n(\mu-\bar x_n) ) ,1-\Phi(\sqrt n(\mu-\bar x_n) ) \}$. See Figure~\ref{pic:cc} for an illustration. The data are generated from $N(0.3,1)$ with sample size 100.
\end{example}

\begin{example}[Example 2] \citep{SinghXieStrawderman2005}
Suppose the data $X_i \sim N(\mu,\sigma^2), i = 1,\ldots, n$, with both unknown $\mu$ and $\sigma$. A \CD for $\mu$ is $H_n (\mu) = F_{t_{n-1}}(\frac{\sqrt n (\mu- \bar x_n )}{s_n})$, where $s_n$ is the sample standard deviation, and $F_{t_{n-1}}(\cdot)$ is the cumulative distribution function of student $t$ distribution with parameter $n-1$. A \CD for $\sigma^2$ is $H_n(\sigma^2) = 1- F_{\chi_{n-1}^2} (\frac{(n-1) s^2_n}{\sigma^2})$, where $F_{\chi_{n-1}^2}(\cdot)$ is the cumulative distribution function of the $\chi^2_{n-1}$-distribution.

\end{example}

\begin{example}[Example 3] \citep{SinghXieStrawderman2005}
Let $\widehat \theta$ be a consistent estimator of $\theta$. For bootstrap, the distribution of $\widehat \theta^* - \theta$ is estimated by the bootstrap distribution $\widehat \theta^* - \widehat \theta$, where $\widehat \theta^*$ is the estimator of $\theta$ computed on a bootstrap sample \citep{EfroTibs93}. An \yifan{asymptotic \CD} for $\theta$ is given by $H_n(\theta) = 1 -\Pr(\widehat \theta^*- \widehat \theta  \leq \widehat \theta  -\theta)= \Pr(\widehat \theta^*\geq 2\widehat \theta  -\theta ) $. In addition, when the limiting distribution of normalized $\widehat \theta$ is symmetric, the raw bootstrap distribution
$H_n(\theta) =  1 -\Pr(\widehat \theta - \widehat \theta^*  \leq \widehat \theta  -\theta) =  \Pr(\widehat \theta^* \leq \theta )$
is also an asymptotic \CD.

\end{example}

\begin{example}[Example 4]
Suppose we are interested in the location parameter $\theta$ of a continuous distribution. When the distribution $F$ is symmetric, i.e., $F(\theta-y)=1-F(\theta+y)$, $\theta$ is the median. The Wilcoxon rank test for $H_0: \theta= t$, $H_1: \theta\neq t$ is based on the summation of signed ranks of $Y_i- t$, i.e., the test statistic $W=\sum_{i=1}^n Z_iR_i$, where $R_i$ is the rank of $|Y_i- t|$, $Z_i$ is an indicator variable with $1$ if $Y_i- t> 0$ and $-1$ otherwise.
Denote by $p(t)$ the \yifan{$p$-value} associated with the Wilcoxon rank test for $H_0: \theta= t$, $H_1: \theta\neq t$. When $t$ varies in $(-\infty, \infty)$, the $p$-value $p(t)$ is referred to as a $p$-value
function. We can prove that the $p$-value
function $p(t)$ is an asymptotic \CD \citep{XieSingh2013}.  Figures~\ref{pic:cc1}
provides illustrations of the asymptotic \CD density $p'(t)$, the asymptotic \CD function $p(t)$ and the asymptotic  CV $2 \min\{p(t), 1-p(t)\}$ for two sample sizes. The data are generated from $N(0,1)$ with sample sizes $n = 10$ and $100$, respectively.

\begin{figure}[h]
\begin{center}
{\includegraphics[width=18cm]{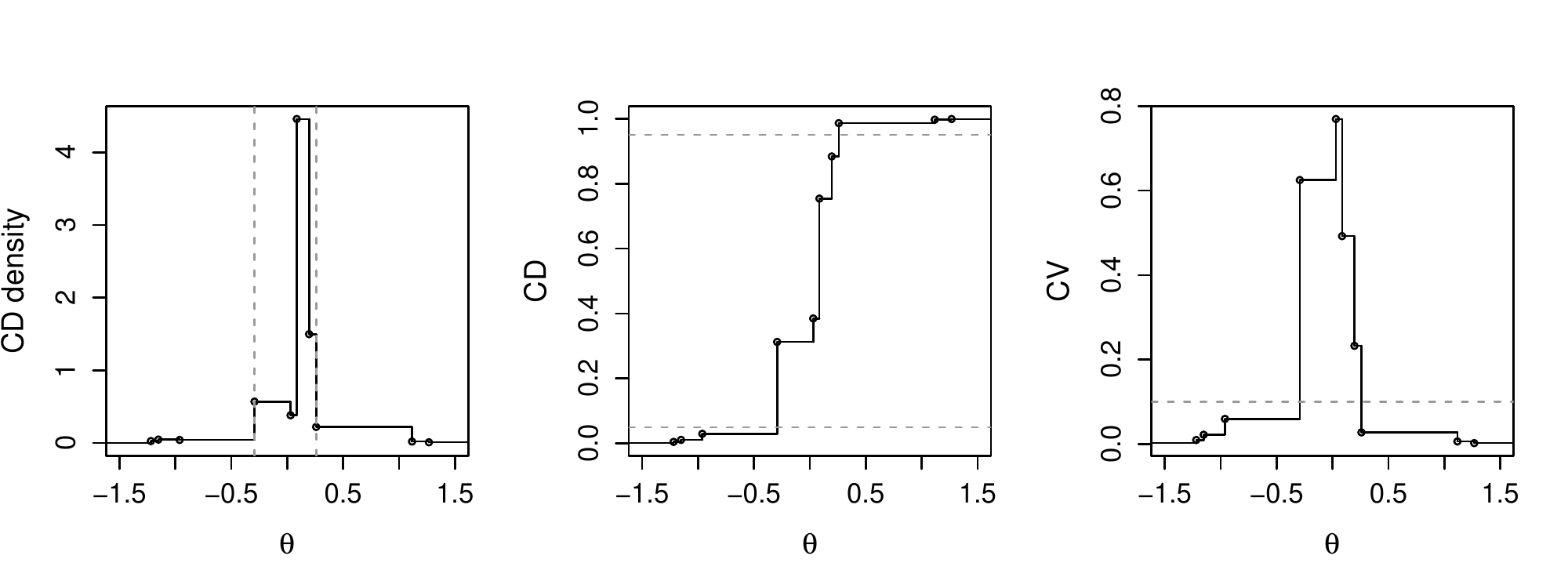}}
{\includegraphics[width=18cm]{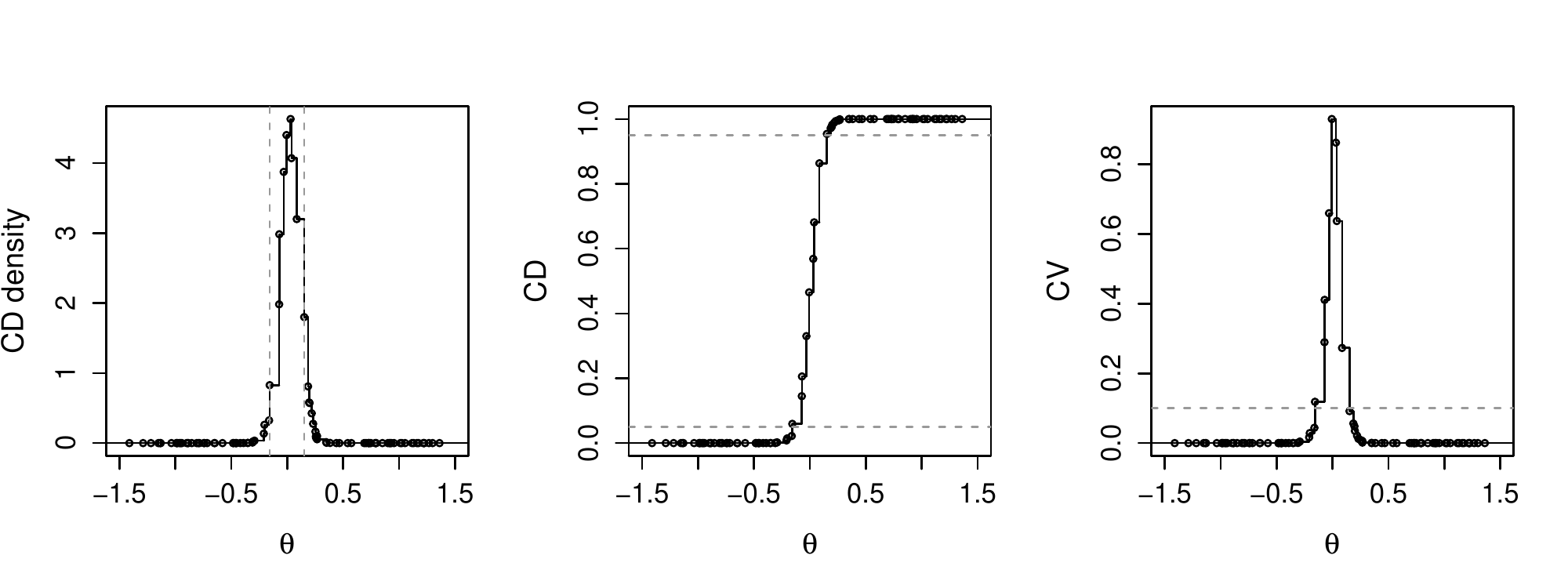}}
\end{center}
\caption{Confidence distributions presented in Example 4 in the forms of density function, cumulative distribution function, and confidence curve. The top row is for sample size $n=10$ and the bottom row is $n=100$.
}
\label{pic:cc1}
\end{figure}


\end{example}

\begin{example}[Example 5]\citep{singhxiestrawderman2007}
Suppose that there is an independent and identically distributed sample of size $n$ from a semi-parametric model involving multiple parameters. Let
$l_n(\theta)$ be the log profile likelihood function and $\mathcal J_n(\theta) = -\ddot l_n(\theta)$  be the observed Fisher information for a scalar parameter of interest $\theta$. Under certain mild assumptions, Theorem 4.1 of \cite{singhxiestrawderman2007} proves that, for any given $\theta$,
\begin{align*}
G_n(\theta) = H_n(\theta) + o_p(1), ~~\text{where}~~ G_n(\theta) = \frac{\int_{-\infty}^\theta \exp\{l_n(x)\}dx }{\int_{-\infty}^\infty \exp\{l_n(x)\}dx }, ~~H_n(\theta) = \Phi \left(  \frac{\theta- \widehat \theta}{\sqrt{\mathcal J_n(\widehat \theta) /n}} \right),~~\widehat \theta = \arg\max_\theta l_n(\theta).
\end{align*}
Because at the true parameter value $\theta = \theta_0$, $H_n(\theta_0)$ converges to $U(0, 1)$ as $n \rightarrow \infty$, it follows that $G_n(\theta_0)$ converges to $U(0, 1)$. Thus, $G_n(\theta)$ is an asymptotic CD.
From this observation, we see that CD-based inference may subsume a likelihood inference in some occasions.
\end{example}
If the sample $X$ is from a discrete distribution, we can typically invoke a large sample theory to obtain an asymptotic CD to ensure the asymptotic frequentist coverage property, when the sample size is large. However,  when the sample size is limited, we sometimes may want to exam the difference between the ``distribution estimator'' and  the $U(0,1)$ distribution to get a sense of under and over coverage. To expand the concept of \CD to cover the cases of discrete distributions with finite sample sizes,
we introduce below the notions of \yifan{lower and upper \CDs}. The lower and upper \CDs  provide us inference statements that are associated with under and over coverages at every significant level.

\begin{definition}[Upper and Lower \CDs]
A function $H_n^+(\cdot)=H_n^+({x},\cdot)$ on $\mathcal X \times \Theta \rightarrow [0,1]$ is said to be an upper \CD for a parameter, if (i) For each given $x \in \mathcal X$, $H_n(\cdot)$ is a monotonic increasing function on $\Theta$ with values ranging within $(0,1)$;  (ii) At the true parameter value $\theta=\theta_0$, $H_n^+(\theta_0) \equiv H_n^+({x},\theta_0)$, as a function of the sample
$x$, is stochastically less than or equal to a uniformly distributed random variable $U \sim U(0, 1)$, i.e.,
\begin{equation}\label{eq:upper}
\Pr\left(H_n^+\big({ X},\theta_0\big) \leq t \right)\geq  t.
\end{equation}
Correspondingly, a lower \CD $H_n^-(\cdot)=H_n(x,\cdot)$ for parameter $\theta$ can be defined but with \eqref{eq:upper} replaced by $ \Pr\left(H_n^-({X},\theta_0) \leq t \right) \leq  t$
for all $t \in (0,1)$.
\end{definition}

More generally, we also refer to  $H_n^+(\cdot)$ and  $H_n^-(\cdot)$ as the upper and lower CD, respectively, even when the monotonic condition (i) is removed. Note that, due to the stochastic dominance inequalities in the definition, we have, for any $\alpha \in (0, 1)$,
\begin{equation*}
\Pr\left(\theta_0 \in \left\{\theta: H_n^+\big({ X}, \theta \big) \leq \alpha \right\} \right)\geq  \alpha\, \hbox{and} \, \Pr\left(\theta_0 \in \left\{\theta: H_n^-({X},\theta) \leq  \alpha \right\}\right) \leq \alpha.
\end{equation*}
Thus, a level-$(1 - \alpha)$  confident interval (or set) $ \left\{\theta: H_n^+\big({ X}, \theta \big) \leq 1 - \alpha \right\}$ or $\left\{\theta: H_n^-({X},\theta) \geq  \alpha \right\}$  has guaranteed the coverage rate of $(1 - \alpha)100\%$, regardless of whether we have the monotonic condition in (i). After we remove the monotonic condition in (i),  $H_n^+(\cdot)$ and  $H_n^-(\cdot)$ may not be a distribution function and the ``nest-ness property'' of confidence intervals/sets may also be lost. Here,  the ``nest-ness property'' refers to
``a level-$(1 - \alpha)$ confidence set C$_{1 - \alpha}$ is not necessarily inside its corresponding  level-$(1 - \alpha')$ confidence set C$_{1 - \alpha'}$, when $1 - \alpha < 1 - \alpha'$''.

To conclude this section, we present an example of lower and upper \CDs.
\begin{example}[Example 6] \citep{HannigXie2012}
Suppose sample $X$ is from Binomial$(n, p_0)$ with observation $x$.
Let $H_n(p,x) = \Pr(X > x) = \sum_{x <k\leq n} {n \choose k} p^k(1-p)^{n-k}$. We can show that $P(H_n(p_0,X) \leq t) \geq t$ and  $P(H_n(p_0,X - 1) \leq t) \leq t$. Thus, $H^+(p,x) = H_n(p,x)$, and $H_n^-(p,x) = H_n(p,x-1)$ are  lower and upper \CDs for the success rate $p_0$.
The half corrected \CD \citep{Efron1998,SchwederHjort2002,Hannig2009} is
\begin{align*}
\frac{H_n^-(p,x)+H_n^+(p,x)}{2} =  \sum_{x<k\leq n_i} {n \choose k} p^k(1-p)^{n-k} +\frac{1}{2} {n \choose x} p^{x}(1-p)^{n-x}.
\end{align*}
\end{example}

\subsection{\CD-based inference \label{sec:inference}}

Analogous to the Bayesian posterior, a \CD contains a wealth of information for constructing any type of frequentist inference.
 We illustrate three aspects of making inference
based on a given \CD. The following Figure~\ref{pic:cd} from \cite{XieSingh2013} provides a graphical illustration of the point estimation, confidence interval, \yifan{hypothesis testing}. More specifically:
\begin{figure}[h]
\begin{center}
\includegraphics[width=13cm]{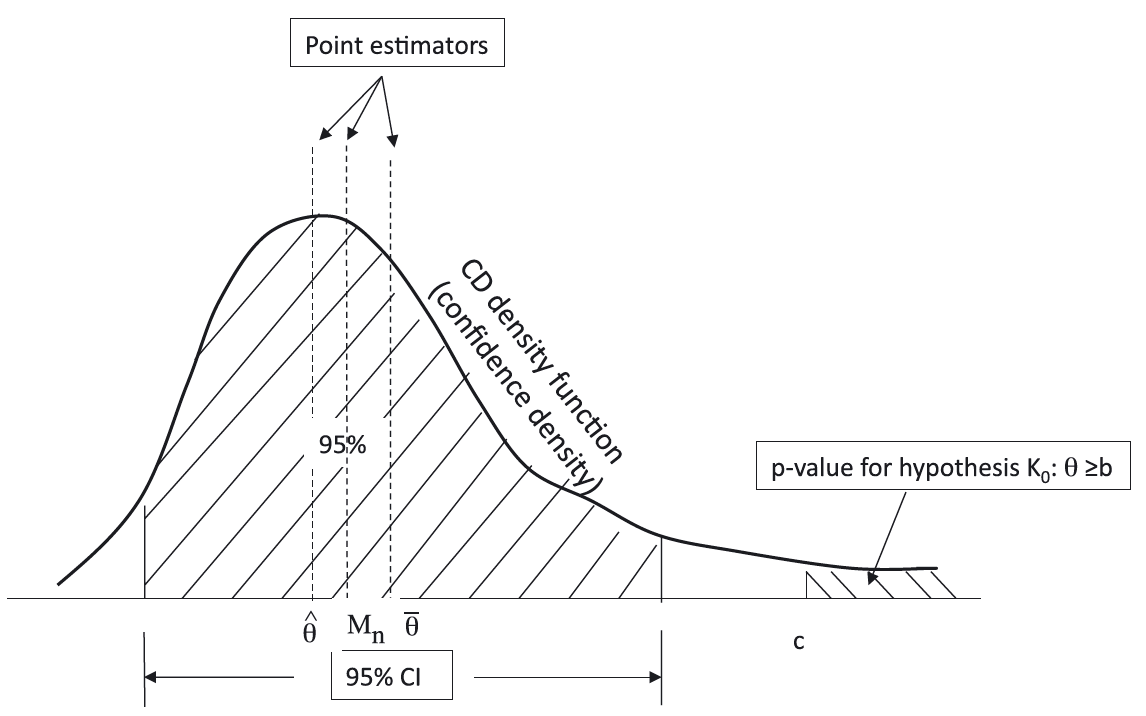}
\end{center}
\caption{A graphical illustration of \CD-based inference \citep{XieSingh2013}}
\label{pic:cd}
\end{figure}

 \textbf{Point estimation}
The natural choices of point estimators of the parameter $\theta$ given a \CD $H_n(\cdot)$, include (i) the median $\widetilde \theta_n=H_n(1/2)$; (ii) the mean $\bar \theta_n=\int_{\theta\in \Theta} \theta d H_n(\theta)$; and (iii) the mode $\widehat \theta_n =\arg \max_{\theta\in \Theta}h_n(\theta)$, where $h_n(\theta)=d H_n(\theta)/d \theta $ is the confidence density function. Under some moderate conditions, these three point estimators are consistent \citep{SinghXieStrawderman2005,XieSinghStrawderman2011,XieSingh2013}.

To further understand these three types of estimators,
the median $\widetilde \theta_n$  is an unbiased estimator with $\Pr_{\theta_0}(\widetilde \theta_n \leq \theta_0)=\Pr_{\theta_0}(1/2 \leq H_n(\theta_0))=1/2$; The mean $\bar \theta_n$ can be viewed as a frequentist analog of Bayesian estimator under the squared loss function; The mode $\widehat \theta_n$ matches with the maximum likelihood estimator if the confidence density is from a normalized likelihood function \citep{XieSingh2013}.

\vspace{0.5cm}

 \textbf{Confidence interval}
As discussed in Section~\ref{sec:cd}, in a confidence curve, a line across the $y$-axis of the significance level $\alpha$  intersects with the confidence curve at two points, and these two points correspond to an $1-\alpha$ level, equal tailed, two-sided confidence interval for $\theta$, i.e., $( H^{-1}_n(\alpha/2), H^{-1}_n(1-\alpha/2) )$.
Furthermore,
$(-\infty, H^{-1}_n(1-\alpha)]$ and $[H^{-1}_n(\alpha),\infty)$ are one-sided $1-\alpha$ level confidence intervals for the parameter $\theta$.

\vspace{0.5cm}

\textbf{Hypothesis testing}
From a \CD, one can obtain $p$-values for various hypothesis testing problems. The natural thinking is to measure the support that $H_n(\cdot)$ lends to a null hypothesis \citep{fraser1991}.
\cite{XieSingh2013} summarized making inference for hypothesis testing from a \CD in the following theorem.
\begin{theorem}
(i) For the one-sided test $K_0:\theta\in C$ versus $K_1:\theta\in C^c$, where $c$ denotes the complementary set, $C$ is an interval of the type of $C_l = (-\infty,b]$ or $C_u = [b,\infty)$, we have $\sup_{\theta\in C} \Pr_{\theta}(p(C) \leq \alpha)=\alpha$ and $p(C)=H_n(C)$ is the corresponding p-value of the test. (ii) For the singleton test $K_0:\theta=b$ versus $K_1: \theta \neq b$, we have $\Pr_{\theta=b}(2\min\{p(C_l),p(C_u)\}\leq \alpha)=\alpha$ and $2 \min\{p(C_l),p(C_u)\}=2\min\{H_n(b),1-H_n(b)\}$ is the p-value of the corresponding test.
\end{theorem}

\begin{example}[Example 6]\citep{XieSingh2013}
Consider Example~2 again. A \CD for $\theta$ is $H_n = F_{t_{n-1}}(\frac{\sqrt n (\mu- \bar x_n )}{s_n})$. For a one-sided test $K_0: \mu \leq b$ versus $K_1 : \mu > b $, its support on the null set $C = (-\infty, b] $ is
\begin{align*}
p(C) = p((-\infty,b] ) =H_n (b) = F_{t_{n-1}}(\sqrt n (b- \bar x_n) /s_n ).
\end{align*}
This is the same $p$-value using the one-sided t-test.
For a two-sided test $K_0: \theta=b$ versus $K_1: \theta \neq b$, the null set $C=\{b\}$. We would like to measure the supports of two alternative sets $p(C^c_{l}) $ and $p(C^c_{u})$.
The rejection region is defined as $\{x : 2\max\{p(C^c_l),p(C^c_u)\}\geq 1-\alpha \}$, i.e.,
\begin{align}
\{x : 2\min\{p(C_l),p(C_u)\}\leq \alpha \} = \{x : 2\min\{H_n(b), 1- H_n(b)\}\leq \alpha \}.
\label{eq:reject}
\end{align}
Under $K_0$ with $\theta = b$, $2 \min \{ p(C_l),p(C_u)   \}= 2 \min \{ H_n(b), 1- H_n(b)   \} \sim U(0,1)$ by the definition of a \CD.
Thus,
\begin{align*}
\text{Pr}_{\theta=b} (2\min\{ p(C_l),p(C_u)  \} \leq \alpha )  = \text{Pr}_{\theta=b} (2\min\{ H_n(b),1- H_n(b)  \} \leq \alpha ) = \alpha
\end{align*}
and the reject region~\eqref{eq:reject} corresponds to a level $\alpha$ test.
Again, the $p$-value $2\min\{p(C_l),p(C_u)\}$ is the standard $p$-value from a two-sided t-test.

\end{example}

\subsection{Combination of \CDs for fusion learning}\label{sec:fusion}

One of the important applications of \CD development is on fusion learning, which synthesizes information from disparate sources with deep implications for meta-analysis \citep{liusingh1997,SchwederHjort2002,SinghXieStrawderman2005,tian2008exact,
XieSinghStrawderman2011, HannigXie2012, XieEtAl2013,liu2014exact,ChenXie2014,claggett2014,Liu2015MultivariateMO,shen2019jasa}.
Fusion learning aims to
combine inference results obtained from different data sources to achieve a more efficient overall inference result.
\CD-based fusion learning applies even when inference results are
derived from different tests or different paradigms, i.e., Bayesian, fiducial, and frequentist~(BFF).

The combination of \CD can be considered as a unified framework for fusion learning.
Suppose there are $k$ independent studies that are dedicated to estimate a common
parameter of interest $\theta$. We assume that we have a \CD $H^i(\cdot)$ for $\theta$ for the sample $x_i$ of the $i$-th study.  \cite{SinghXieStrawderman2005} proposed a
general recipe for combining these $k$ independent \CDs:
\begin{align}
H^{c}(\theta) \equiv G_c \{ g_c(H^1(\theta), \ldots, H^k(\theta) ) \},
\label{eq:fl1}
\end{align}
where
$g_c$ is a given continuous function on $[0, 1]$ which is non-decreasing in each coordinate, the function $G_c$ is determined by the monotonic function $g_c$ with $G_c(t) = \Pr(g_c(U_1, . . . ,U_k)\leq t)$, and $U_1,\dots,U_k$ are independent uniform random variables. The function $H_c(\cdot)$ contains information from all $k$ samples and is referred to as a combined \CD for the parameter $\theta$.
Furthermore, the \CD obtained by Equation~\eqref{eq:fl1} does not require any information regarding how the input \CDs are obtained.

A special class of the general combining framework~\eqref{eq:fl1} plays a prominent role
in unifying many modern \yifan{meta-analysis} approaches. The choice of the
function $g_c$ for this special class is
\begin{align}
g_c(u_1,\ldots, u_k)= w_1 F^{-1}(u_1) + \cdots + w_k F^{-1}(u_k),
\label{eq:fl2}
\end{align}
where $F(\cdot)$ is a given cumulative distribution function, and $w_i\geq 0$ with at least one $w_i \neq 0$ are generic weights for the combination rule. Generally, there are two types of weights:
fixed weights to improve the efficiency of combination and adaptive weights based
on data.

As shown in \cite{XieSinghStrawderman2011}, it is remarkable that by choosing different $g_c$ functions, all the classic approaches of combining $p$-values including Fisher, Normal (Stouffer), Min (Tippett), Max, and Sum methods \citep{Marden1991} and all the five model-based meta-analysis estimators described in \cite{Normand1999} including the maximum likelihood method and Bayesian approach under fixed-effects model; method of moment estimators, restricted maximum likelihood method, and Bayesian estimator with a normal prior under random-effects model, can all be obtained through a \CD combination framework.
Furthermore, it was shown in \cite{yang2016} that Mantel-Haenszel and Peto methods, as well as Tian et al.'s method of combining confidence intervals  \citep{tian2008exact} for meta-analysis of $2 \times 2$ tables can  also all be obtained through a \CD combination framework.
An R-package ``gmeta'' developed by \cite{yang2017} implements the \CD combining framework for fusion learning including
classical $p$-value combination methods from \cite{Marden1991}, meta-analysis estimators with both fixed-effects and random-effects models, and many other approaches.

Fusion learning under the framework of combining \CD provides an extensive and powerful tool for synthesizing information from diverse data sources. This approach has sound theoretical support and has been applied to many practical situations
including robust fusion learning \citep{XieSinghStrawderman2011}, exact fusion learning for discrete data \citep{tian2008exact,liu2014exact}, fusion learning for heterogeneous studies \citep{Liu2015MultivariateMO}, non-parametric fusion learning \citep{liusingh1997,claggett2014}, split-conquer-combine approach \citep{ChenXie2014}, and individualized fusion learning ($\textit{i}$-Fusion) \citep{shen2019jasa}, etc. We refer to \cite{cheng2017} for more detailed discussions.

\subsection{Multivariate \CDs}

A simultaneous \CD for vector parameters can sometimes be difficult to define \citep{SchwederHjort2002}, especially on how to define a multivariate \CD in the exact sense in some non-Gaussian settings 
to ensure that their marginal distributions are \CDs for the corresponding single parameter.
We consider the Behrens-Fisher problem  of testing for the equality of means from two multivariate normal distributions when the covariance matrices are unknown and possibly not equal. A joint \CD of the two population means $(\mu_1,\mu_2)$ has a joint density of the form
 $$f_1\left(\frac{\mu_1 -\bar x_1}{s_1/\sqrt n_1}\right)f_2\left(\frac{\mu_2-\bar x_2}{s_2/\sqrt n_2}\right)/\left(s_1s_2\sqrt{n_1n_2}\right),$$
 where $f_i$ is the density function for the student t-distribution with $n_i-1$ degrees of freedom, $i=1,2$.
 The marginal distribution of $\mu_1-\mu_2$ is only an asymptotic \CD but not a \CD in the
exact sense.

The good news in the multi-dimensional case is that under asymptotic settings or wherever bootstrap theory applies, one can still work with multivariate \CDs \citep{XieSingh2013}. When no analytic confidence curve for the parameter vector $\theta$ of interest is available,
the product method of \cite{beran1988} can be used if confidence curves are available
for each component of the vector \citep{SchwederHjort2002}. Additionally, if we only consider center-outwards confidence regions instead of all Borel sets in the $p\times 1$ parameter space, the central-\CDs considered in \cite{singhxiestrawderman2007} and the confidence net considered in \cite{Schweder2007} offer coherent notions of multivariate \CDs in the exact sense \citep{XieSingh2013}.

There are many approaches to obtain \CDs. One way is normalizing
a likelihood function curve with respect to its parameters so that the area underneath the curve is one. The normalized likelihood
function is typically a density function. For instance, under some mild conditions, \cite{Fraser1984} show that this normalized likelihood function is the normal density function of an asymptotic \CD. Other ways like bootstrap distributions and $p$-value functions also often provide valid \CDs.
Finally,  \CDs and fiducial distributions have been always linked since their inception. The class of fiducial inference provides another systematic way to obtain \CDs and we will further discuss fiducial inference in the next section.

\section{Fiducial inference}

\CD can be somehow viewed as ``the Neymanian interpretation of Fisher's fiducial distributions'' \citep{schweder2016confidence}. From the definition of \CD and fiducial distribution, we may consider the fiducial distribution as one special type of \CD, though the \CD looks at the problem of obtaining an inferentially meaningful distribution on the parameter space from a pure frequentist point of view \citep{XieSingh2013}. Nevertheless, fiducial inference provides a systematic way to obtain a \CD, and its development provides a rich class of literature for \CD inference. We briefly review fiducial inference and its recent developments in this section.

\subsection{Fiducial inference}
R. A. Fisher introduced the idea of fiducial probability and fiducial inference \citep{Fisher1930}
as a potential replacement of the Bayesian posterior distribution. Although he discussed fiducial inference in several subsequent papers, there appears to be no rigorous definition of a fiducial distribution for a vector parameter.
The basic idea of the fiducial argument is switching the role of data and parameters to introduce the distribution on the parameter space. This obtained distribution then summarizes our knowledge about the unknown parameter.
Since the mid 2000s, there has been a renewed interest in modern modifications of fiducial inference.
The common approaches for these modifications rely on a definition of inferentially meaningful probability statements about subsets of the parameter space without introducing any prior information.

These modern approaches include generalized fiducial inference \citep{Hannig2009,hannig2016generalized}, Dempster-Shafer theory \citep{Dempster2008, EdlefsenLiuDempster2009}, inferential models \citep{martin2015inferential,ryan2017IM}.  Objective Bayesian inference, which aims at finding non-subjective model-based priors, can also be seen as addressing the same question. Examples of recent breakthroughs related to reference prior and model selection are  \citet{BayarriEtAl2012, BergerBernardoSun2009, BergerBernardoSun2012}.
Another related approach is based on higher-order likelihood expansions and implied data-dependent priors
\citep{FraserFraserStaicu2010,
Fraser2004,
Fraser2011,
FraserNaderi2008,
Fraser:2005tc,
FraserReidMarrasYi2010}. There are many more references that interested
readers can find in \cite{hannig2016generalized}.

\subsection{Generalized fiducial distribution}

Generalized fiducial inference,  motivated by \cite{TsuiWeerahandi1989, TsuiWeerahandi1991},  has been at the forefront of the modern fiducial revival.
Generalized fiducial inference defines a data-dependent measure on the parameter space by using an inverse of a deterministic data generating equation without the use of Bayes theorem.

Motivated by Fisher's fiducial argument, generalized fiducial inference begins with expressing the relationship between the data $\bY$ and the parameters $\btheta$ as
\begin{equation}\label{eq:StructuralEq}
 \bY = \bG(\bU,\btheta),
\end{equation}
where $\bG(\cdot,\cdot)$ is a deterministic function termed as the data generating equation, and $\bU$ is the random component of this data generating equation whose distribution is independent of parameters and completely known.

The data $\bY$ are created by generating a random variable $\bU$ and plugging it into the data generating equation \eqref{eq:StructuralEq}.
For example, a single observation from $N(\mu,1)$ distribution can be
written as $Y=\mu+U$, where $\btheta=\mu$ and $U$ is $N(0,1)$ random variable.

Fisher's original fiducial argument only addresses the simple case where the data generating equation~\eqref{eq:StructuralEq} can be inverted and the
inverse $Q_\by(\bu)=\btheta$ exists for any observed
$\by$ and for any arbitrary $\bu$.
One can define the fiducial distribution for $\btheta$ as the distribution of
$Q_\by(\bU^\star)$ where $\bU^\star$ is an independent copy of
$\bU$.  Equivalently, a sample from the fiducial distribution of
$\btheta$ can be obtained by first generating $\bU^\star_i,$ and then let $\btheta_i^\star=Q_\by(\bU^\star_i)$, $i=1,\ldots,n$. Point estimation and confidence intervals for $\btheta$ can be obtained based on this sample. In the $N(\mu,1)$
example, $Q_y(u)=y-u$ and the fiducial distribution is therefore the distribution of $y-U^\star\sim N(y,1)$.

In the case of no $\theta$ satisfying Equation~\eqref{eq:StructuralEq}, \cite{Hannig2009} proposed to use the distribution of $U$ conditional on the event $\{u : y =G(u, \theta),~ \text{for some}~ \theta \}$.
\cite{hannig2016generalized} generalized this approach and proposed an attractive definition of generalized fiducial distribution (GFD) through a weak limit.
\begin{definition}\label{d:FiducialDefinition}
A {probability measure} on the parameter space $\bTheta$ is called a GFD if it can be obtained as a weak limit
 \begin{equation}\label{eq:FiducialLimit}
  { \lim_{\epsilon\to 0} \left[\argmin_{\btheta^\star} \|\by-\bG(\bU^\star,\btheta^\star)\| \ \Big|\ \min_{\btheta^\star}\|\by-\bG(\bU^\star,\btheta^\star)\|\leq \epsilon\right]}.
 \end{equation}
 \end{definition}

\cite{hannig2016generalized} pointed out a close relationship between \GFD and Approximate Bayesian {Computations} (ABC) \citep{BeaumontEtAl2002}.  In an idealized ABC, one first generates an observation {$\btheta^*$} from the prior, then generates a new sample using a data generating equation $\by^\star=\bG(\bU^\star,\btheta^\star)$ and compares the generated data with the observed data $\by$. If the observed and generated data sets are close, i.e., $\|\by-\by^\star\|\leq\epsilon$, the generated $\btheta^\star$ is accepted, otherwise it is rejected and the procedure is repeated.
 On the other hand,  as for \GFD, one first generates $\bU^\star$, finds a best fitting $\btheta^\star=\argmin_{\btheta^\star} \|\by-\bG(\bU^\star,\btheta^\star)\|$, computes $\by^\star=\bG(\bU^\star,\btheta^\star)$, again accepts $\btheta^\star$ if $\|\by-\by^\star\|\leq\epsilon$ and rejects otherwise. In either approach an artificial data set $\by^\star=  \bG(\bU^\star,\btheta^\star)$ is generated and compared to the observed data. The main difference is that the Bayes posterior simulates the parameter $\btheta^\star$ from the prior while \GFD uses the best fitting parameter.

Fiducial distributions often have good frequentist properties and corresponding fiducial confidence intervals often give asymptotically correct coverage \citep{Hannig2009,hannig2016generalized}. In addition, fiducial distribution is a data-dependent measure on the parameter space and thereby a \CD.
\cite{XieSingh2013} described the relation between the concepts of \CD and fiducial distributions using an analogy in point estimation:
A \CD is analogous to a consistent estimator and a fiducial distribution is analogous to a maximum likelihood estimator.
In the context of point estimation, a consistent estimator does not have to be a maximum likelihood estimator. But under some regularity conditions, the maximum likelihood estimator typically provides a standard procedure to obtain a consistent estimator.
In the context of distribution estimator, a \CD does not have to be a fiducial distribution. However, under suitable conditions, a
fiducial distribution often has good frequentist properties and thus a \CD.

\subsection{A user friendly formula for GFD}

While Definition~\eqref{eq:FiducialLimit} for \GFD is conceptually and mathematically appealing, it is not clear how to compute the limit in most of practical situations.
The following theorem proposed by \cite{hannig2016generalized} provides a computational tool.

\begin{theorem}\label{t:FiducialFormula}
Under certain assumptions, the limiting distribution in \eqref{eq:FiducialLimit} has a density
\begin{equation}\label{eq:GreatFiducial}
    r(\btheta|\by)=\frac{f(\by,\btheta) J(\by,\btheta)}{\int_\bTheta f(\by,\btheta') J(\by,\btheta')\,d\btheta'},
\end{equation}
where $f(\by,\btheta)$ is the likelihood and the function
\begin{equation}\label{eq:RecommendedJacobian}
 J(\by,\btheta)=D\left(\left.\frac{\bd}{\bd\btheta} \bG(\bu,\btheta)\right|_{\bu=\bG^{-1}(\by,\btheta)}\right).
\end{equation}
If (i) {$n=p$} then {$D(A)=|\det A|$}. Otherwise the function $D(A)$ depends on the norm used;
(ii) the $l_\infty$ norm gives {$D(A)=\sum\limits_{\mathbf i=(i_1,\ldots,i_p)}\left|
    {\det(A)_{\mathbf i}} \right|$}; \footnote{In (ii) the sum spans over $\binom np$ of $p$-tuples of indexes $\bi=(1\leq i_1<\cdots< i_p\leq n)$. For any $n\times p$ matrix $A$, the sub-matrix $(A)_\bi$ is the $p\times p$ matrix containing the rows $\bi=(i_1,\ldots,i_p)$ of $A$.}
(iii) under an additional assumption stated in \cite{hannig2016generalized}, the $l_2$ norm gives  {$D(A)=(\det A^\top A)^{1/2}$}.

\end{theorem}

 \cite{hannig2016generalized} recommended using~(ii) for practitioners. A nice property of \GFD is that \GFD is invariant under smooth re-parameterizations.
This property follows directly from \eqref{eq:FiducialLimit}, since for an appropriate selection of minimizers and any one-to-one function  $\btheta=\phi(\etab)$,
\[\phi\left(\argmin_{\etab^\star} \|\by-\bG(\bU^\star,\phi(\etab^\star))\|\right)=\argmin_{\btheta^\star} \|\by-\bG(\bU^\star,\btheta^\star)\|.\]

Note that \GFD could change with transformations of the data generating equation.
Assume that the observed data set has been transformed by a one-to-one smooth transformation
$
\bZ= \bT(\bY).
$
By the chain rule, the \GFD based on this new data generating equation and observed data $\bz=\bT(\by)$ is the density \eqref{eq:GreatFiducial} with the Jacobian function
\begin{equation}\label{eq:RecommendedJacobian3}
 J_\bT(\bz,\btheta)=
D\left(\left.\frac{\bd}{\bd\by} \bT(\by) \cdot \frac{\bd}{\bd\btheta} \bG(\bu,\btheta)\right|_{\bu=\bG^{-1}(\by,\btheta)}\right),
\end{equation}
where for simplicity we write $\by$ instead of $T^{-1}(\bz)$.

\subsection{Examples of \GFD}

In this section we will consider two examples, linear regression and uniform distribution. In the first case the \GFD is the same as Bayes posterior with respect to the independence Jeffreys prior while in the second case, the \GFD is not a Bayes posterior with respect to any prior (that is not data dependent).

\begin{example}[Linear Regression \citep{hannig2016generalized}]
We consider a generalized fiducial approach to regression problem.
We express linear regression via the data generating equation,
\[
 \bY = G(\bU, \btheta)=\bX \beta +\sigma \bU,
\]
where $\bY$ is the dependent variables, $\bX$ is the design matrix, $\btheta=(\beta,\sigma)$ are the unknown parameters and $\bU$ is a random vector with known density $f(\bu)$ independent of $\theta$ and $X$.
Note that
$\frac{\bd}{\bd\btheta} \bG(\bU,\btheta) = (\bX,\bU)$ and $\bU=(\by-\bX\beta)/\sigma$,
the Jacobian in \eqref{eq:RecommendedJacobian3} using the $l_\infty$ norm simplifies to
\[
J_\infty(\by,\btheta)=  \sigma^{-1} \sum_{\substack{\bi=(i_1,\ldots,i_p) \\ 1\leq i_1<\cdots<i_p\leq n}}\left|\det\left(\bX, \bY\right)_\bi\right|,
\]
and the density of \GFD is
\[
 r(\beta,\sigma | \by)\propto \sigma^{-n-1} f((\bY-\bX\beta)/\sigma).
\]
The fiducial solution is the same as the Bayesian solution using Jeffreys prior \citep{Berger2011}.
Furthermore, by a simple calculation, the Jacobian with $l_2$ norm differs from $J_\infty(\by,\btheta)$ only by a constant, the \GFD remains unchanged.

\end{example}

\begin{example}[GFD in irregular models \citep{hannig2016generalized}]\label{ex:Uniform}

We consider an irregular model $U\big(a(\theta)-b(\theta),a(\theta)+b(\theta)\big)$.
The reference prior for this model has been shown complex in Theorem~8 from \cite{BergerBernardoSun2009}. Consider \GFD approach, we first express the observed data by the following data generating equation
\[
 Y_i=a(\theta)+b(\theta) U_i,\quad U_i\  \overset{i.i.d.}{\sim} \  U(-1,1).
\]
By simple algebra,
\begin{align*}
\frac{d}{d\theta} G(\bu,\theta) = a'(\theta)+b'(\theta)\bU \quad \text{with} \quad  \bU=b^{-1}(\theta)(Y-a(\theta)).
\end{align*}
If $a'(\theta)>|b'(\theta)|$, \eqref{eq:RecommendedJacobian} simplifies to
\begin{equation*}\label{eq:jocobunif1}
     J_1(\by,\theta)
                    =n[a'(\theta)-a(\theta)\{\log b(\theta)\}'+\bar y_n\{\log b(\theta)\}'],
\end{equation*}
and the \GFD is
\begin{equation*}\label{eq:fidunif}
 r_1(\theta|\by)\propto \frac{a'(\theta)-a(\theta)\{\log b(\theta)\}'+\bar y_n\{\log b(\theta)\}'}{b(\theta)^n}I_{\{a(\theta)-b(\theta)<y_{(1)}\ \&\ a(\theta)+b(\theta)>y_{(n)}\}}.
\end{equation*}

Consider an alternative fiducial solution, which constructs the \GFD based on the minimal sufficient and ancillary statistics $Z=\{h_1(Y_{(1)}),h_2(Y_{(n)}),(Y-Y_{(1)})/(Y_{(n)}-Y_{(1)})\}^\top$, where $Y_{(1)}, Y_{(n)}$ are order statistics,
$  h_1^{-1}(\theta)=EY_{(1)}=a(\theta)-b(\theta)(n-1)/(n+1) \mbox{ and } h_2^{-1}(\theta)=EY_{(n)}=a(\theta)+b(\theta)(n-1)/(n+1).$
 By a simple calculation,
\begin{align*}\label{eq:jocobunif2}
 J_2(\by,\theta)&=(w_1+w_2)\left[a'(\theta)-a(\theta)\{\log b(\theta)\}'+\frac{w_1 y_{(1)}+w_2 y_{(n)}}{w_1+w_2}\{\log b(\theta)\}'\right],\\
 r_2(\theta|\by)& \propto \frac{I_{\{a(\theta)-b(\theta)<y_{(1)}\ \&\ a(\theta)+b(\theta)>y_{(n)}\}}  }{\left[(w_1+w_2)[a'(\theta)-a(\theta)\{\log b(\theta)\}']+ (w_1 y_{(1)}+w_2 y_{(n)})\{\log b(\theta)\}'\right]^{-1} b(\theta)^n},
 \end{align*}
where $w_1=h_1'(y_{(1)})$ and $w_2=h_2'(y_{(n)})$.

\cite{hannig2016generalized} performed extensive simulation studies for a particular case $U(\theta,\theta^2)$ comparing \GFD to the Bayesian posteriors with the reference prior $\pi(\theta)=\frac{(2\theta-1)}{\theta(\theta-1)}e^{\psi\left(\frac{2\theta}{2\theta-1}\right)}$ \citep{BergerBernardoSun2009}\footnote{$\psi(x)$ is the digamma function defined by $\psi(z)=\frac{d}{dz}\log(\Gamma(z))$ for $z > 0$, where $\Gamma$ is Gamma function.} and flat prior $\pi(\theta)=1$. The simple \GFD, the alternative \GFD, and the reference prior Bayes posterior maintain nominal coverage for all parameter settings. However, the flat prior Bayes posterior does not have a satisfactory coverage, with the worst departures from nominal coverage for small sample size and large parameter $\theta$.

\end{example}

\begin{example}[\yifan{Nonparametric fiducial inference} with right censored data \citep{Cui2019}]

Let failure times $X_i~(i=1,\ldots,n)$ follow the true  distribution function $F_0$ and censoring times $C_i ~(i=1,\ldots,n)$ have the distribution function $R_0$. We treat the situation when failure and censoring times are independent and unknown. Suppose we observe right censored data $\{y_i,\delta_i\}$ $(i=1,\ldots n)$, where $y_i=x_i\wedge c_i$ is the minimum of $x_i$ and $c_i$, $\delta_i=I\{x_i\leq c_i\}$ denotes censoring indicator.

Consider the following data generating equation,
\begin{align*}
Y_i=F^{-1}(U_i)\wedge R^{-1}(V_i),\quad \Delta_i=I\{F^{-1}(U_i)\leq R^{-1}(V_i)\} \quad (i = 1,\ldots n),
\end{align*}
where $U_i, V_i$ are independent and identically distributed $U(0,1)$.

For a failure event $\delta_i=1$, we have full information about failure time $x_i$, i.e., $x_i=y_i$, and partial information about censoring time $c_i$, i.e., $c_i\geq y_i$.
Thus, $$F^{-1}(u_i)=y_i   \Longleftrightarrow F(y_i)\geq u_i, F(y_i-\epsilon)< u_i ~\text{for any}~ \epsilon>0.$$

For a censored event $\delta_i=0$, we only know partial information about $x_i$, i.e., $x_i > y_i$, and full information on $c_i$, i.e., $c_i= y_i$. Similarly,
\begin{align*}
F^{-1}(u_i)> y_i &\Longleftrightarrow F(y_i)< u_i,\\
R^{-1}(v_i)=y_i &\Longleftrightarrow R(y_i)\geq v_i, R(y_i-\epsilon)< v_i ~\text{for any}~ \epsilon>0.
\end{align*}

The complete inverse map of the data generating equation is
\begin{equation}\label{eq:QP2}
Q^{F,R}(\by,\bdelta,\bu,\bv)=\bigcap_i Q^{F,R}_{\delta_i}(y_i,u_i,v_i)=
Q^{F}(\by,\bdelta,\bu)\times Q^{R}(\by,\bdelta,\bv),
\end{equation}
where
\begin{equation}\label{eq:QP2F}
Q^F(\by,\bdelta,\bu)=\left\{F:
\begin{cases}
  F(y_i)\geq u_i,  F(y_i-\epsilon)< u_i ~\text{for any}~ \epsilon>0 & \mbox{for all  $i$ such that $\delta_i=1$}\\
  F(y_j)< u_j                         &  \mbox{for all  $j$ such that $\delta_j=0$}
 \end{cases}
  \right\} ,
\end{equation}
and $Q^R(\by,\bdelta,\bv)$ is analogous.

Let $(\bU^*,\bV^*)$ be an independent copy of $(\bU,\bV)$.
Because the inverse \eqref{eq:QP2} separates into a Cartesian product, and the fact that  $\bU^*$ and $\bV^*$ are independent, the marginal fiducial distribution for the failure distribution function $F$ is
\begin{equation*}\label{eq:GFDsurv}
Q^F(\by,\bdelta,\bU^*) \mid \{Q^F(\by,\bdelta,\bU^*) \neq \emptyset\}.
\end{equation*}

Figure~\ref{explain} from \cite{Cui2019} demonstrates the survival function representation of $Q^F(\by,\bdelta,\bu)$, as defined in Equation~\eqref{eq:QP2F},  for one data set with $n=8$ observations of  $X$ following $Weibull(20,10)$ censored by $Z$ following $Exp(20)$. Each of the panels corresponds to a different value of $\bu$, where each $u$ is a realization of $U^*$. Any survival function lying between the upper red and the lower black fiducial survival functions corresponds to an element of the closure of $Q^F(\by,\bdelta,\bu)$.
The technical details of sampling refer to Algorithm~1 in \cite{Cui2019}. The corresponding fiducial-based confidence intervals proposed in \cite{Cui2019} maintain coverage in situations where asymptotic methods
often have substantial coverage problems. Furthermore, as also shown in \cite{Cui2019}, the average length of their log-interpolation fiducial confidence intervals is often shorter than the length of confidence intervals for competing methods that maintain coverage.
As pointed by \cite{Taraldsen2019discuss}, it would also be interesting to consider other choices of fiducial samples such as monotonic spline interpolation.

\begin{figure}[H]\centering
\includegraphics[width=80mm]{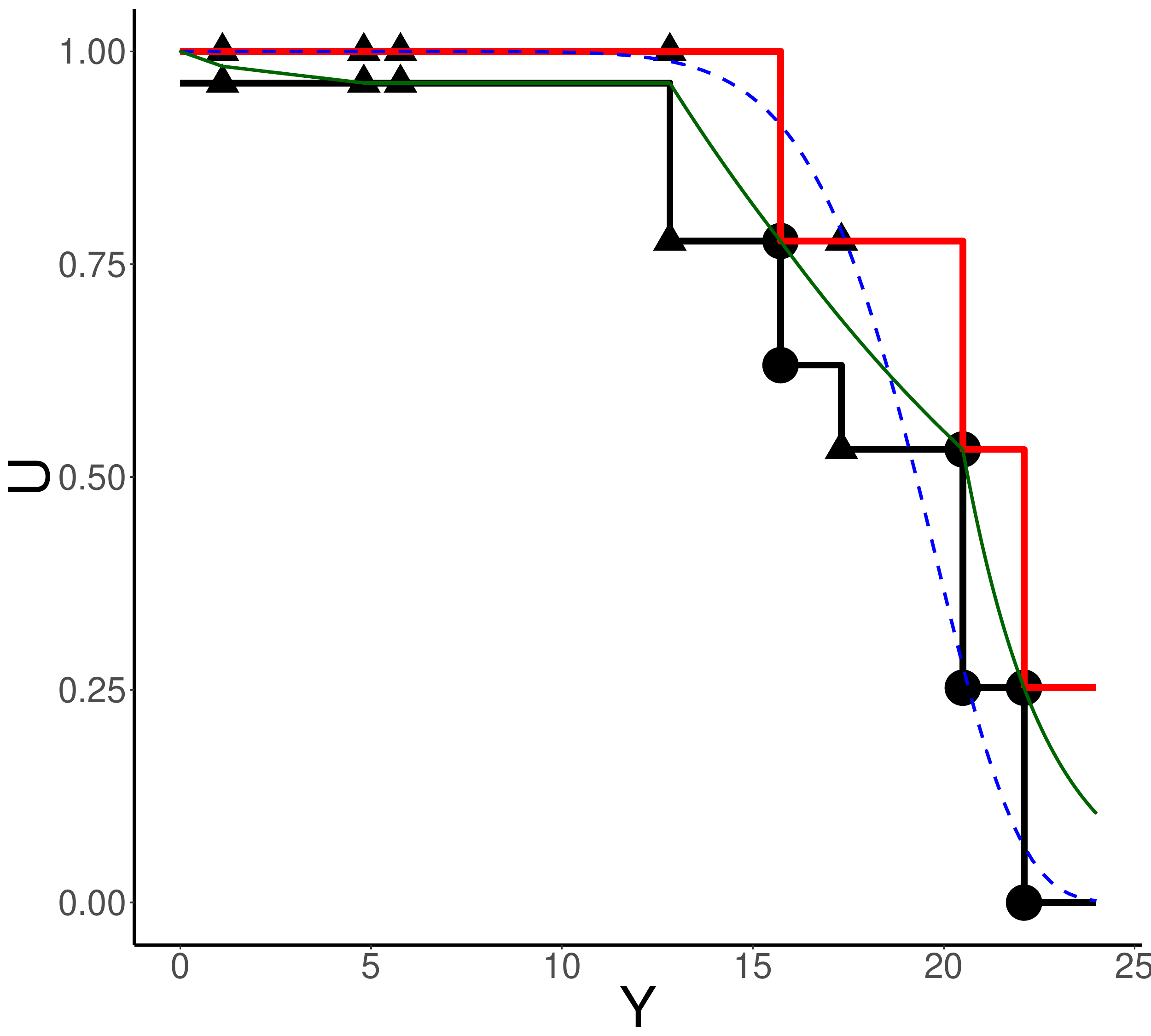}
\includegraphics[width=80mm]{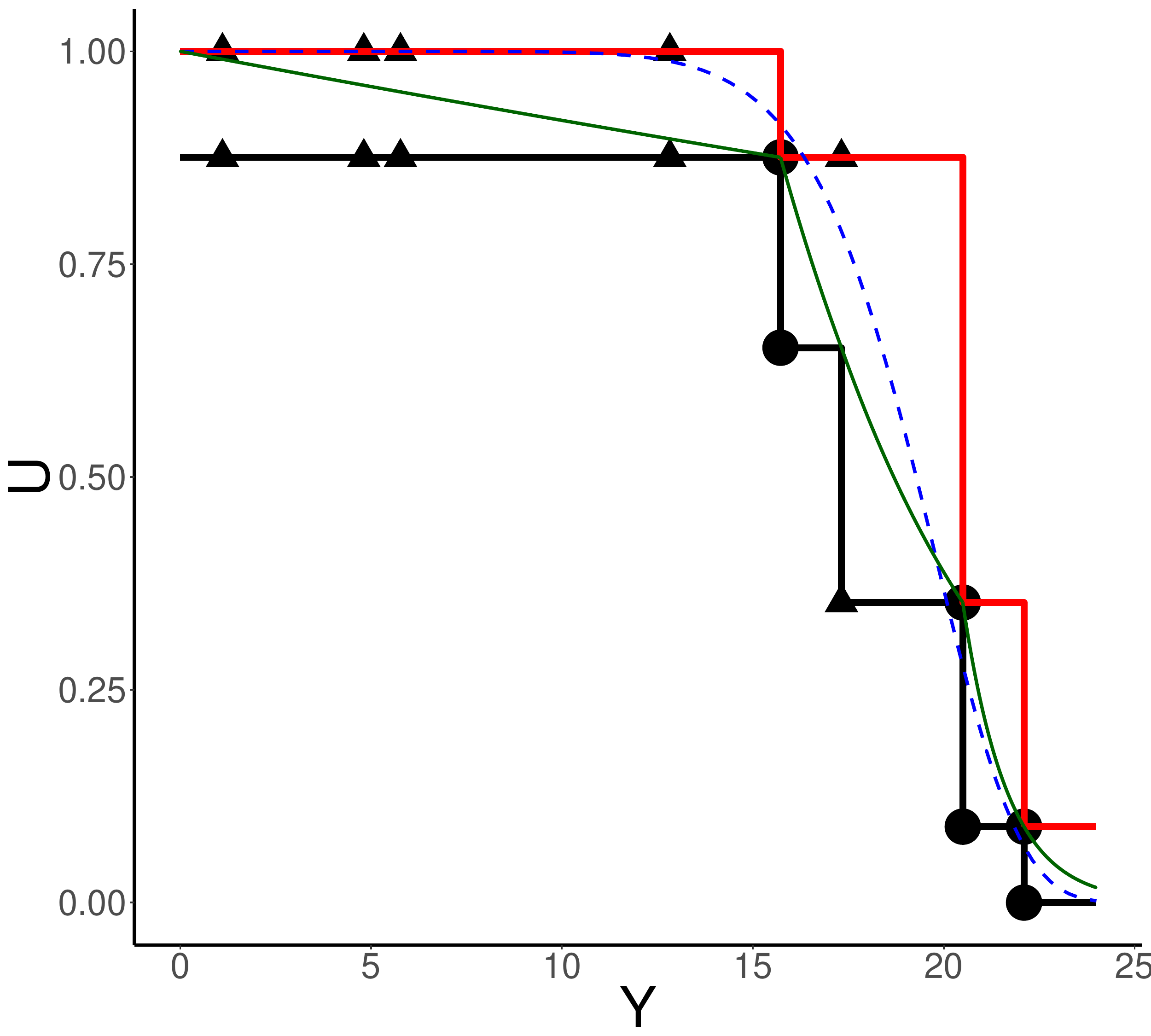}
\caption{Two realizations of fiducial curves for a sample of size $8$ from $Weibull(20,10)$ censored by $Exp(20)$ \citep{Cui2019}. Here fiducial curves refer to \yifan{Monte Carlo samples} $S^L_i$, $S^U_i$, and $S^I_i$ ($i=1,2$) from the \GFD. The red and black curves are corresponding realizations of the upper and lower fiducial survival functions. The green curve is the log-linear interpolation type of survival functions. The circle points denote failure observations. The triangle points denote censored observations. The dashed blue curve is the true survival function of $Weibull(20,10)$.}
\label{explain}
\end{figure}
\end{example}

\begin{example}[GFDs for discrete distributions \citep{hannig2016generalized}]
Let $Y$ be a random variable with distribution function $F(y|\theta)$. Assume there is $\mathcal Y$ so that $P_\theta(Y\in\mathcal Y)=1$ for all $\theta$, and for each fixed $y\in \mathcal Y$ the distribution function is either a non-increasing function of $\theta$, spanning the whole interval $(0,1)$, or a constant equal to $1$; the left limit $F(y_-|\theta)$ is also either a non-increasing function of $\theta$ spanning the whole interval $(0,1)$, or a constant equal to $0$.

Define $F^-(a|\theta)=\inf\{y: F(y|\theta)\geq a\}$. It is well known \citep{CasellaBerger2002} that if $U\sim$ U(0,1),
$
 Y=F^-(U|\theta)
$
has the correct distribution and we use this association as a data generating equation.
It follows that both $Q^+_y(u)=\sup\{\theta: F(y|\theta)=u\}$ and $Q^-_y(u)=\inf\{\theta: F(y_-|\theta)=u\}$ exist and satisfy $F(y|Q^+_y(u))=u$ and $F(y_-|Q^-_y(u))=u$.
Consequently,
 \[
  P(Q^+_y(u)\leq t)=1-F(y|t) \quad\mbox{and}\quad P(Q^-_y(u)\leq t)=1-F(y_-|t).
 \]
Note that for all $u\in(0,1)$, the function $F^-(u|\theta)$ is non-increasing in $\theta$  and the closure of the inverse image $\bar{Q}_y(u)=\{Q^-_y(u),Q^+_y(u)\}$. 
The half corrected \GFD has distribution function
\[
 R(\theta| y)=1-\frac{F(y|\theta)+F(y_-|\theta)}2.
\]
If either of the distribution functions is constant we interpret it as a point mass at the appropriate boundary of the parameter space.
Analogous argument shows that if the distribution function and its left limit were non-decreasing in $\theta$, the half corrected \GFD would have distribution function
\[
 R(\theta| y)=\frac{F(y|\theta)+F(y_-|\theta)}2.
\]
\cite{hannig2016generalized} provide a list of the half corrected GFDs for three well known discrete distributions. Let Beta$(0,n+1)$ and Beta$(x+1,0)$ denote the degenerate distributions on $0$ and $1$, respectively. Let $\Gamma(0,1)$ denote the degenerate distribution on $0$.
\begin{itemize}
 \item $X\sim$ Binomial$(m,p)$ with $m$ known. \GFD is the mixture of {Beta}$(x+1,m-x)$ and {Beta}$(x,m-x+1)$ distributions \citep{Hannig2009}.
 \item $X\sim$ Poisson$(\lambda)$.  \GFD is the mixture of $\Gamma(x+1,1)$ and $\Gamma(x,1)$ distributions \citep{Dempster2008}.
 \item $X\sim$ Negative Binomial$(r,p)$ with $r$ known. \GFD is the mixture of {Beta}$(r,x-r+1)$ and {Beta}$(r,x-r)$ distributions \citep{Hannig2014}.
\end{itemize}
\end{example}

\begin{example}[Model Selection via GFD \citep{hannig2016generalized}]
\cite{HannigLee2009} introduced model selection into the generalized fiducial inference paradigm in the context of wavelet regression. Two important ingredients are needed for fiducial model selection:  1) include the choice of model as one of the parameters; 2) include penalization in the data generating equation.

Consider a finite collection of models $\mathcal M$. The data generating equation is
\begin{equation}\label{eq:dgamodel}
 \bY=\bG(M, \btheta_M,\bU),\qquad M\in\mathcal M,\ \btheta_M\in\bTheta_M,
\end{equation}
where $\bY$ is the observation, $M$ is the model considered, $\btheta_M$ includes the parameters associated with model $M$, and $\bU$ is a random vector of with fully known distribution independent of any parameters.
\cite{HannigLee2009} proposed a novel way of adding a penalty into the fiducial model selection. In particular, for each model $M$, they proposed to augment the data generating equation \eqref{eq:dgamodel} by
\begin{equation}\label{eq:dgeaugment}
 0=P_k,\quad k=1,\ldots,\min(|M|,n),
\end{equation}
where $P_k$ are independent and identically distributed continuous random variables independent of $\bU$ with $f_P(0)=q$, and $q$ is a constant determined by the penalty. \cite{HannigLee2009} recommended using $q=n^{-1/2}$ as the default penalty. Note that the number of additional equations is the same as the number of unknown parameters in the model. As we never actually observe the outcomes of the extra data generating equations, we will select their values as $p_i=0$.

For the augmented data generating equation we have the following theorem from \cite{hannig2016generalized}. The quantity $r(M|\by)$ can be used for inference in the usual way. For example, fiducial factor, the ratio $r(M_1|\by)/r(M_2|\by)$, can be used in the same way as a Bayes factor, as discussed in \cite{BergerPericchi2001} in the context of Bayesian model selection.
\begin{theorem}[\cite{hannig2016generalized}]\label{t:modelselection}
Suppose $|M|\leq n$ and certain assumptions hold, the marginal generalized fiducial probability of model $M$ is
\begin{equation}
   r(M|\by)=\frac{q^{|M|} \int_{\bTheta_M} f_M(\by,\btheta_M) J_M(\by,\btheta_M)\,d\btheta_M}{\sum_{M'\in\mathcal M}q^{|M'|}\int_{\bTheta_{M'}} f_{M'}(\by,\btheta_{M'}) J_{M'}(\by,\btheta_{M'})\,d\btheta_{M'}},
\label{eq:rmodelselect}
\end{equation}
where $f_M(\by,\btheta_M)$ is the likelihood and $J_M(\by,\btheta_M)$ is the Jacobian function computed using \eqref{eq:RecommendedJacobian3} for each fixed model $M$.
\end{theorem}
For more details on the use of fiducial model selection, see \cite{HannigLee2009} and \cite{LaiHannigLee2014}.
\end{example}

\section{Applications and Numerical examples}

\subsection{\CD based inference}

\textbf{Two-parameter exponential distribution}
Inference procedures based on the two-parameter exponential model, $Exp(\mu, \sigma)$, are extensively used in several areas of statistical practice, including survival and reliability analysis. The probability distribution function and cumulative distribution function of a random variable $X \sim Exp(\mu,\sigma)$ are given, respectively, by
\begin{align*}
f(x)&=\frac{1}{\sigma}\exp\{-\frac{x-\mu}{\sigma}\},\\
F(x)&=\begin{cases}
      1-\exp\{-\frac{x-\mu}{\sigma}\} & \text{if}~ x>\mu,\\
      0 &  \text{if}~ x\leq \mu,
    \end{cases}
\end{align*}
and survival function (also known as reliability function) is $S(x)=1-F(x)$.
The inference problem of interest is to obtain confidence intervals (sets) of $\mu$, $\sigma$ and $S(t)$ at a given $t > 0$.

Let $X_{(1)},\ldots, X_{(k)}$ be the $k$ ($k>1$) smallest observations among $X_{1},\ldots, X_{n}$. Then the maximum likelihood estimator of $\mu$ and $\sigma$ are
\begin{align*}
\widehat \mu= X_{(1)},\quad \quad \text{and} \quad \quad
\widehat \sigma= \frac{1}{k} \left\{ \sum_{i=1} ^k X_{(i)}+(n-k)X_{(k)}- nX_{(1)} \right\}.
\end{align*}
It turns out that $\widehat \mu$ and $\widehat \sigma$ are independent and they follow the distributions,
\begin{align}
\label{eq:exp}
U=2n(\widehat \mu-\mu )/\sigma \sim \chi^2(2), \quad
V=2k\widehat \sigma/\sigma \sim \chi^2(2k-2),
\end{align}
respectively. Here $\chi^2(m)$ is the Chi-square distribution with degree of freedom $m$.
We provide below a simple \CD-based method to answer the inference problem of interest.

From Equation~\eqref{eq:exp}, we have
$$
\frac{n(\widehat \mu-\mu)}{k \widehat \sigma}  = \frac{U/2}{V/(2k-2)} \sim F(2, 2k-2),
$$
where $F(a, b)$ is the $F$-distribution with degrees of freedom $a$ and $b$.
By the pivot-based \CD construction method \citep[][p134]{singhxiestrawderman2007}, a \CD for $\mu$ is $H_1(\mu) = 1 - F_{F(2, 2k-2)}(\frac{n(\widehat \mu-\mu)}{k \widehat \sigma})$, where  $F_{F(2, 2k-2)}$ is the cumulative distribution function of $F(2, 2k-2)$-distribution. Similarly, a \CD for $\sigma$ is $H_2(\sigma) = 1 - F_{\chi^2(2k-2)}(\frac{2k(\widehat \sigma)}{\sigma})$, where  $F_{\chi^2(2k-2)}$ is the cumulative distribution function of $\chi^2(2k-2)$-distribution.
Inferential statements regarding $\mu$ and $\sigma$, including confidence intervals and testing results, can be obtained from these two \CDs. Coverage rates and test errors obtained from these two \CDs are exact.

We can also consider the inference for $(\mu, \sigma)$ jointly. Here, we introduce a simulation-based approach. Let $U^* \sim  \chi^2(2)$ and $V^* \sim \chi^2(2k-2)$ be two independently simulated random numbers. Define
$$
\xi^* =  \widehat \mu - \frac{k\widehat \sigma }{n} \frac{U^* }{V^* }   \quad \hbox{and} \quad \zeta^*  = \frac{2k \widehat \sigma}{V^*}.
$$
Then, $\xi^{*} | (\widehat \mu, \widehat \sigma) \sim H_1(\mu)$ and $\zeta^{*} | (\widehat \mu, \widehat \sigma) \sim H_2(\sigma)$, and they are called {\it \CD random variables} \citep{XieSingh2013}. Furthermore, the underlying joint distribution of $(\xi^*, \zeta^* )$, given $(\widehat \mu, \widehat \sigma)$, is a joint \CD function $H_3(\mu, \sigma)$ of $(\mu, \sigma)$. If we simulate a large number of, say $M$,  copies of $(U^*, V^*)$, then we can get $M$ copies of $(\xi^*, \zeta^*)$. In order to make inference statements about $(\mu, \sigma)$, we can treat these $M$ copies of $(\xi_1^*, \zeta_1^*), \ldots, (\xi_M^*, \zeta_M^*)$ as if they were $M$ copies of bootstrap estimators in bootstrap inference, or as if they were $M$ copies of random samples from the posterior distribution of $(\mu, \sigma)$ in a Bayesian inference.

Additionally, we can also use the $M$ copies of \CD random variables $(\xi_1^*, \zeta_1^*), \ldots, (\xi_M^*, \zeta_M^*)$ to obtain a pointwise confidence band for $S(t)$, $t > 0$. For each given $t >0$, we compute $\kappa_j^*(t) =  \exp\{ - (t - \xi_j^*)/\zeta_j^*\}$, for $j = 1, \ldots, M$. Then $[\kappa_{[\alpha M]}^*(t), +\infty)$ and $[\kappa_{[\frac{\alpha}2 M]}^*(t), \kappa_{[\frac{(1 - \alpha)}{2} M]}^*(t)]$ are the one-sided and two-sided level-$\alpha$ confidence intervals of $S(t)$, respectively, where $\kappa_{[qM]}^*(t)$ is the $q$-th quantile of $\kappa_1^*(t), \ldots, \kappa_M^*(t)$.  Now by varying $t$,
$[\kappa_{[\alpha M]}^*(t),+\infty)$ forms a level-$\alpha$ lower confidence band and $[\kappa_{[\frac{\alpha}2 M]}^*(t), \kappa_{[\frac{(1 - \alpha)}{2} M]}^*(t)]$ forms a level-$\alpha$ confidence band for the survival function $S(t)$.

\begin{figure}[h]
\begin{center}
\includegraphics[width=7cm]{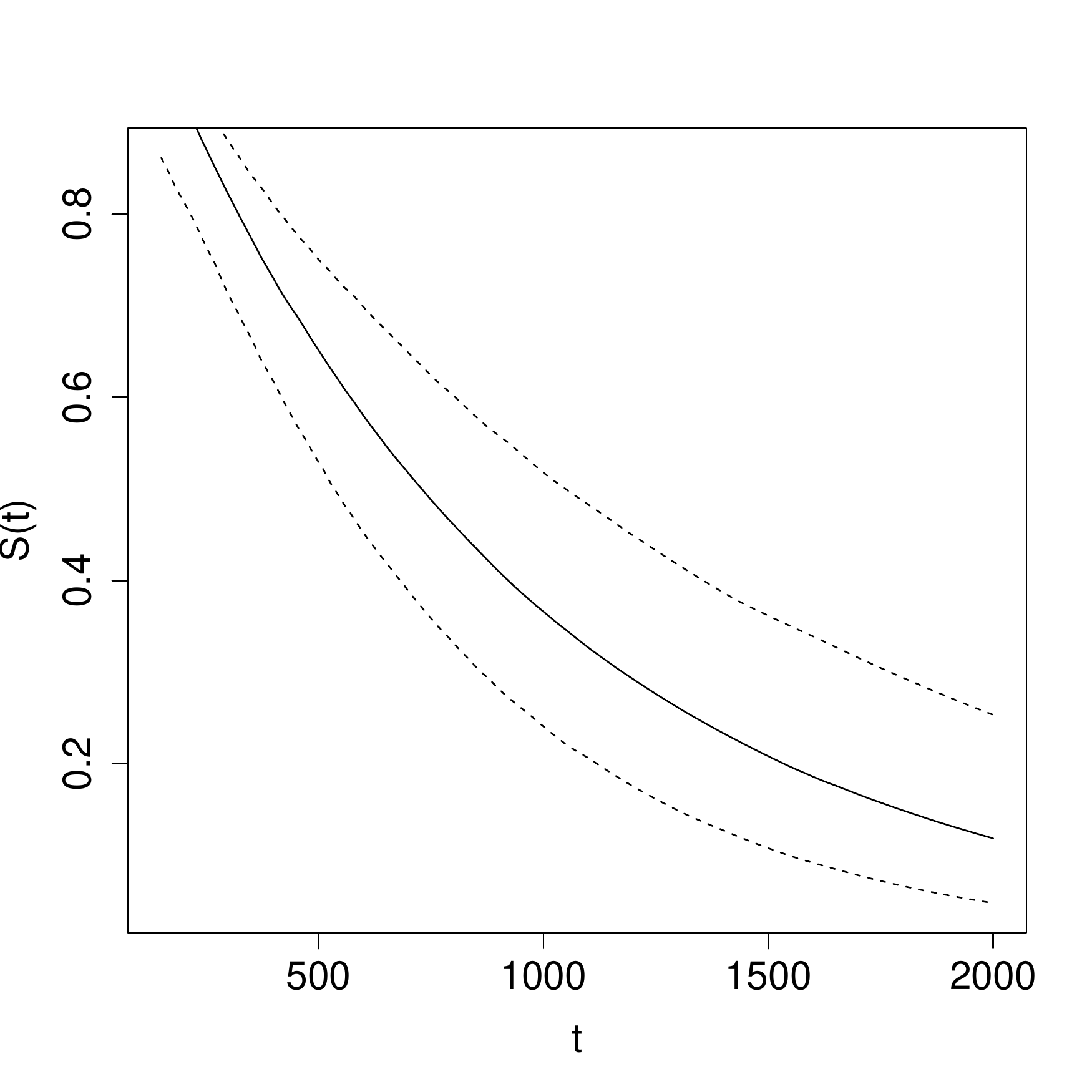}
\end{center}
\caption{Point estimate (solid line) and 95\% confidence band (dashed line) of \CD-based inference}
\label{pic:gcl}
\end{figure}

We can show that this set of \yifan{exact} confidence bands derived from the \CD method matches with those obtained in \cite{roy2005} using Tsui and Weerahandi's \yifan{generalized inference approach} \citep{TsuiWeerahandi1989}, but  the \CD approach is very simple and more direct.  \cite{roy2005} illustrated the 95\% lower limit $\widetilde S(t)$ for time ranging from 150 to 2000 in Figure~1 of \cite{roy2005} using a real data example with 19 observations taken from \cite{Lawless1982}. The data deal with mileages for military personnel carriers that failed in service. Figure~\ref{pic:gcl} is a similar plot for the confidence band, using our \CD approach with $M=1000$.

\vspace{0.5cm}

Data \citep{Lawless1982}:

162, 200, 271, 320, 393, 508, 539, 629, 706, 777, 884, 1008, 1101, 1182, 1463, 1603, 1984, 2355, 2880

\vspace{0.5cm}

\begin{figure}[h]
\begin{center}
\includegraphics[width=5.5cm]{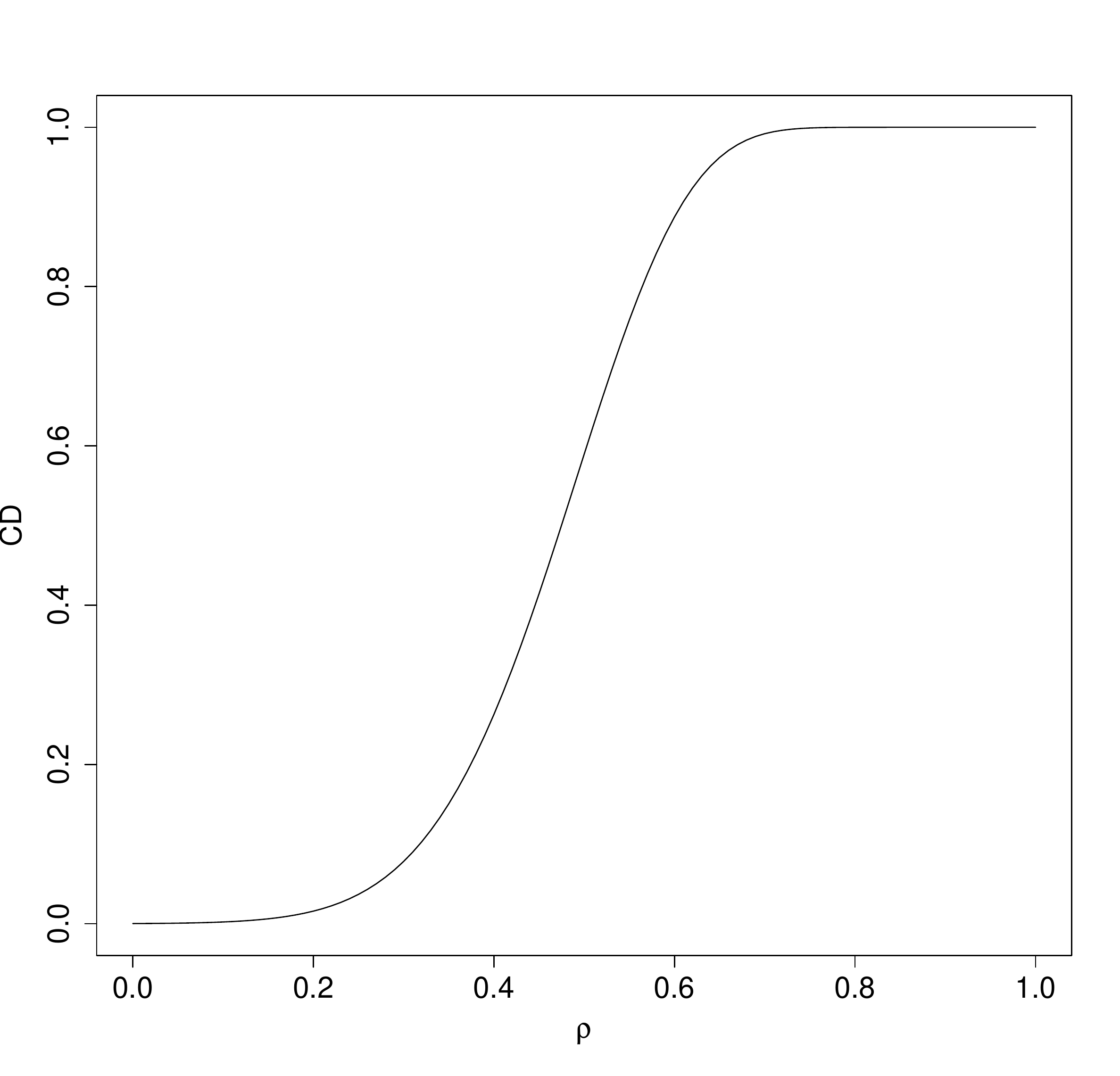}
\end{center}
\caption{CD of the correlation coefficient $\rho$}
\label{pic:bivariate}
\end{figure}

\textbf{Bivariate normal correlation}
Suppose we have the following bivariate normal distribution,
\[
N\left(\begin{pmatrix}
\mu_1\\
\mu_2
\end{pmatrix},\begin{pmatrix}
\sigma_1^2 & \rho\sigma_1\sigma_2 \\
\rho\sigma_1\sigma_2 & \sigma_2^2
\end{pmatrix}\right),
\]
and let $\rho$ denote the correlation coefficient. One could use the asymptotic pivot, Fisher's $Z$ \citep{fisher1915,singhxiestrawderman2007},
\begin{align*}
\frac{1}{2}\log\frac{1+r}{1-r}-\frac{1}{2} \log \frac{1+\rho}{1-\rho},
\end{align*}
where $r$ is the sample correlation.  The limiting distribution of the above pivot is $N(0,\frac{1}{n-3})$.
Therefore, the asymptotic \CD is
\begin{align*}
H_n(\rho) = 1- \Phi\left( \sqrt{n-3}  \left[\frac{1}{2}\log\frac{1+r}{1-r}-\frac{1}{2} \log \frac{1+\rho}{1-\rho}\right] \right), ~~ -1\leq \theta\leq 1.
\end{align*}
Figure~\ref{pic:bivariate} presents the CD of correlation coefficient $\rho$ for a simulated dataset with $n=50, \mu_1=\mu_2=1, \sigma_1=\sigma_2=1, \rho=0.5$.

In addition to the above two examples, there also are recent developments of CDs on causal inference, see more applications in \cite{luo2020leveraging}.

\subsection{Nonparametric GFD based inference}
\cite{Cui2019} proposed a fiducial approach to testing reliability function with an infinite dimensional parameter. Their approach does not assume a parametric distribution and is robust to model mis-specification. In \cite{Cui2019}, they considered a clinical trial of chemotherapy against chemotherapy combined with radiotherapy in the treatment of locally unresectable gastric cancer conducted by the Gastrointestinal Tumor Study Group \citep{schein1982comparison}. In this trial, forty-five patients were randomized to each of the two groups and followed for several years. The censoring percentage is 13.3\% for the combined therapy group, and 4.4\% for the chemotherapy group. We are interested in testing whether the two treatment groups have the same survival functions.

\begin{figure}[h]\centering
\begin{subfigure}[b]{0.40\textwidth}
\centering
\includegraphics[height=48mm]{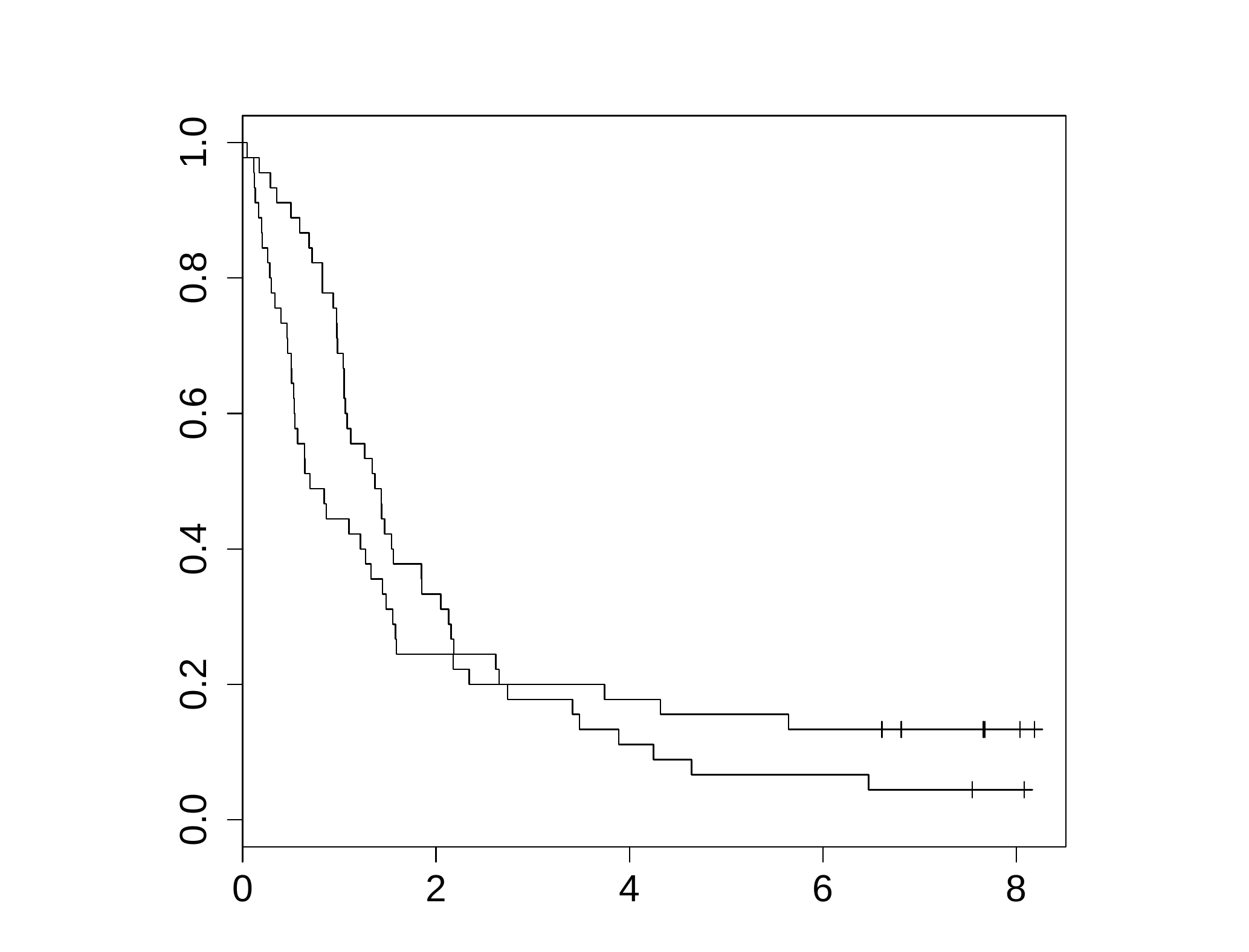}
\caption{Kaplan-Meier estimators for two treatment groups \citep{Cui2019}.}
\label{KM}
\end{subfigure}
\hspace*{0.5cm}
\begin{subfigure}[b]{0.40\textwidth}
\centering
\includegraphics[height=45mm]{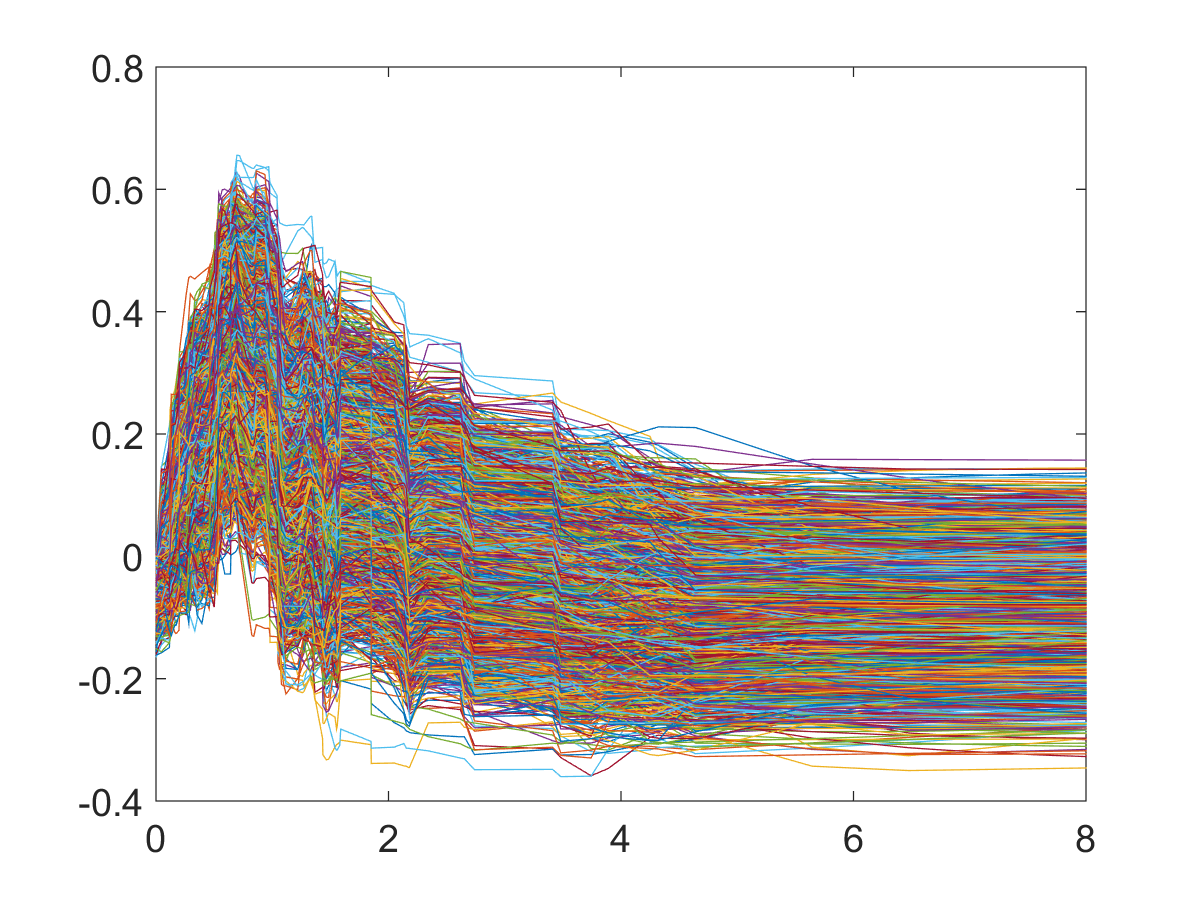}
\caption{Difference of two sample fiducial distributions. 
}
\label{realdiff}
\end{subfigure}
\caption{}
\end{figure}

The Kaplan-Meier curves for these two datasets are presented in Figure~\ref{KM}. We notice that the two hazards appear to be crossing, which could pose a problem for some log-rank tests. In this instance, the fiducial approach gives a small p-value 0.002. The p-values of other types of log-rank tests are reported in \cite{Cui2019}. To explain why their proposed fiducial approach works good, they plot the sample of the difference of two fiducial distributions
in Figure~\ref{realdiff}. If these two datasets are from the same distribution, 0 should be well within the sample curves. However, from Figure~\ref{realdiff}, we could see that the majority of curves are very far
away from 0 on the interval [0.5, 1]. This gives strong evidence that the group with combined therapy has significantly worse early survival outcomes.

In \cite{Cui2019}, they choose to use the sup-norm in the definition of the curvewise confidence intervals and tests. It could be possible to make the procedure more powerful by using a different (possibly weighted) norm \citep{nair1984confidence}.
Similarly, it might also be possible to use the choice of norm motivated by inferential models \citep{martin2015inferential,martin2019discuss,cuihannig2019rejoinder}.
Besides the above example, there also are recent developments of nonparametric fiducial inference on interval censored data and Efron's empirical Bayes deconvolution, see  \cite{cui2020fiducial,cui2020survival} for more applications.

\vspace{0.5cm}

Data \citep{schein1982comparison}: (* indicates censored event)

Combination group: 0.05 0.12 0.12 0.13 0.16 0.20 0.20 0.26 0.28 0.30 0.33 0.39 0.46 0.47 0.50 0.51 0.53 0.53 0.54 0.57 0.64 0.64 0.70 0.84 0.86 1.10 1.22 1.27 1.33 1.45 1.48 1.55 1.58 1.59 2.18 2.34 3.74 4.32 5.64 6.61* 6.81* 7.66* 7.68* 8.04* 8.19*

Chemotherapy group: 0.00 0.17 0.29 0.35 0.50 0.59 0.68 0.72 0.82 0.82 0.94 0.97 0.98 0.98 1.04 1.05 1.05 1.06 1.08 1.12 1.26 1.34 1.37 1.43 1.44 1.47 1.54 1.56 1.85 1.85 2.05 2.13 2.15 2.18 2.62 2.65 2.74 3.41 3.48 3.89 4.25 4.64 6.47 7.55* 8.08*

\subsection{Combining information from multiple \CDs}

We use simple cluster of differentiation 4 (cd-4)  count data considered in \cite{diciccio1996} to demonstrate combining information from \CDs. Twenty HIV-positive subjects received an experimental antiviral drug. The cd-4 counts in hundreds were recorded for each subject
at baseline and after one year of treatment.

We obtained the summary statistics and simulated four independent datasets from the following bivariate normal distribution
\[
N\left(\begin{pmatrix}
\mu_1\\
\mu_2
\end{pmatrix},\begin{pmatrix}
\sigma_1^2 & \rho\sigma_1\sigma_2 \\
\rho\sigma_1\sigma_2 & \sigma_2^2
\end{pmatrix}\right),
\]
where $\mu_1=3.288$, $\mu_2=4.093$, $\sigma_1^2=0.657$, $\sigma_2^2= 1.346$, and $\rho=0.723$.

Suppose each study makes its own inference conclusion individually.
Each dataset was analyzed by  Fisher's $Z$ method \citep{fisher1915,DescTools}, the bias-corrected and accelerated ($\text{BC}_a$) bootstrap \citep{diciccio1996,Davison1997,Canty2019}, the profile likelihood approach \citep{li2018profileR}, and Bayesian with uniform prior \citep{baath2013}, respectively. One natural question we would like to ask is if we can combine the inferences from four independent studies, given that $\rho$ is the same in all studies. The answer is yes. As introduced in Section~\ref{sec:fusion}, combination of \CDs is a powerful inferential tool. We fused studies by combining $p$-values (Stouffer) \citep{Marden1991,yang2017r}.

\begin{table}[!h]
\center
\caption{Inference on correlation coefficient: combining independent bivariate normal studies}
\begin{tabular}{ccc}
 \hline
Methods & 95\% CI  & CDs \\
 \hline
Fisher's $Z$ method & (0.348,0.845)  &  \raisebox{-.5\height}{\includegraphics[width=4.2cm]{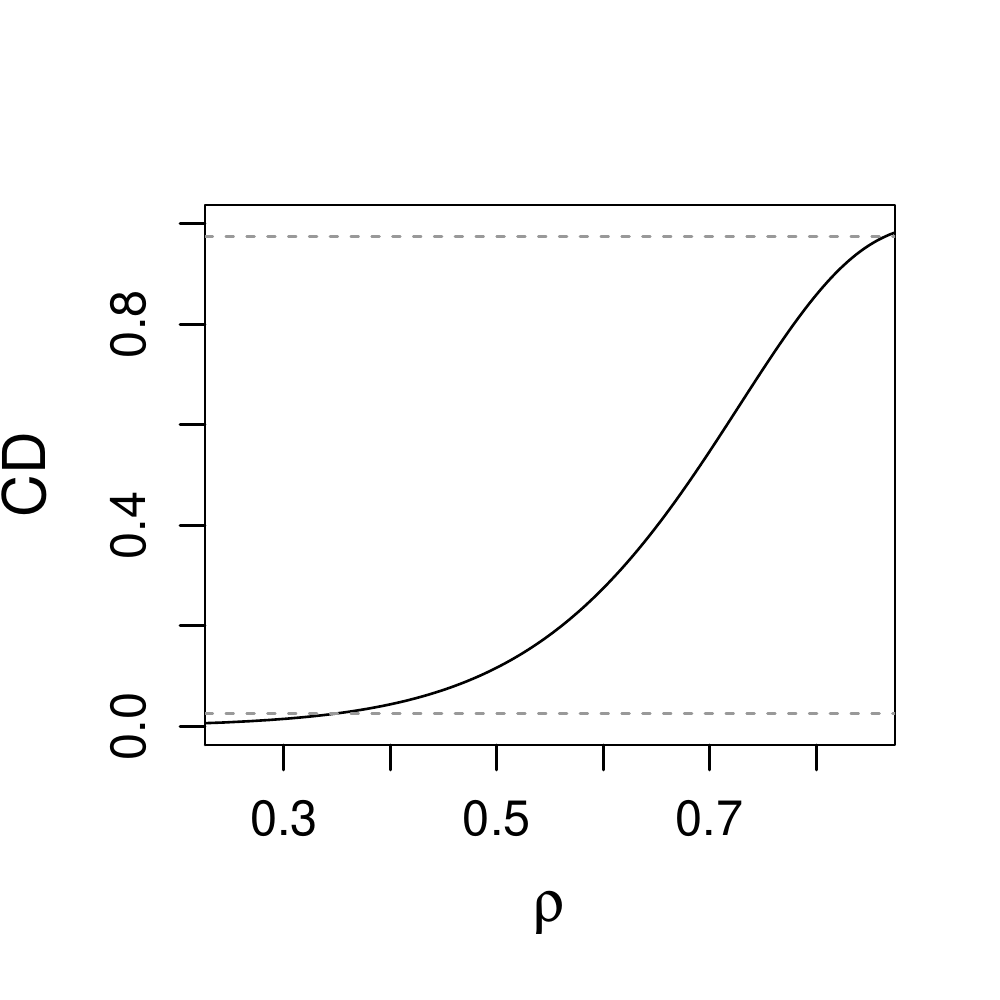}} \\
Bootstrap $\text{BC}_a$ & (0.317,0.818) & \raisebox{-.5\height}{\includegraphics[width=4.2cm]{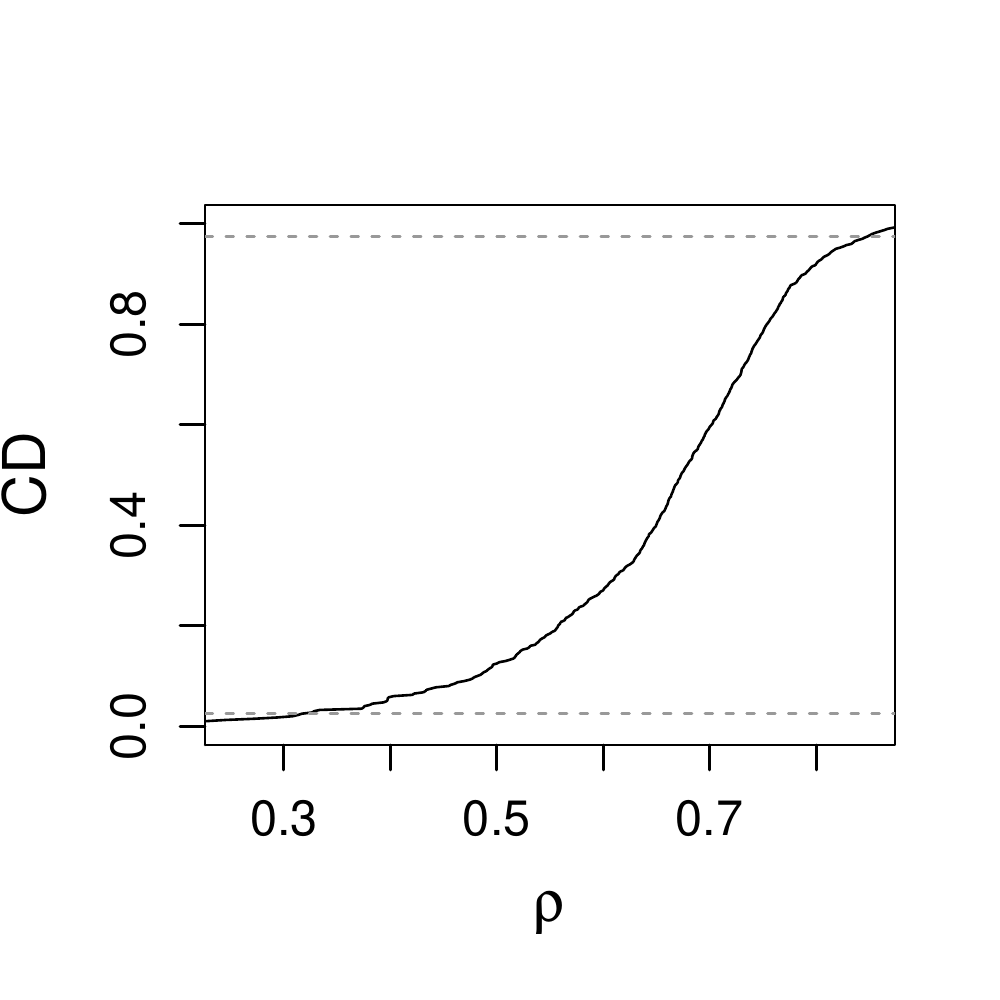}} \\
Profile likelihood & (0.346,0.827) & \raisebox{-.5\height}{\includegraphics[width=4.2cm]{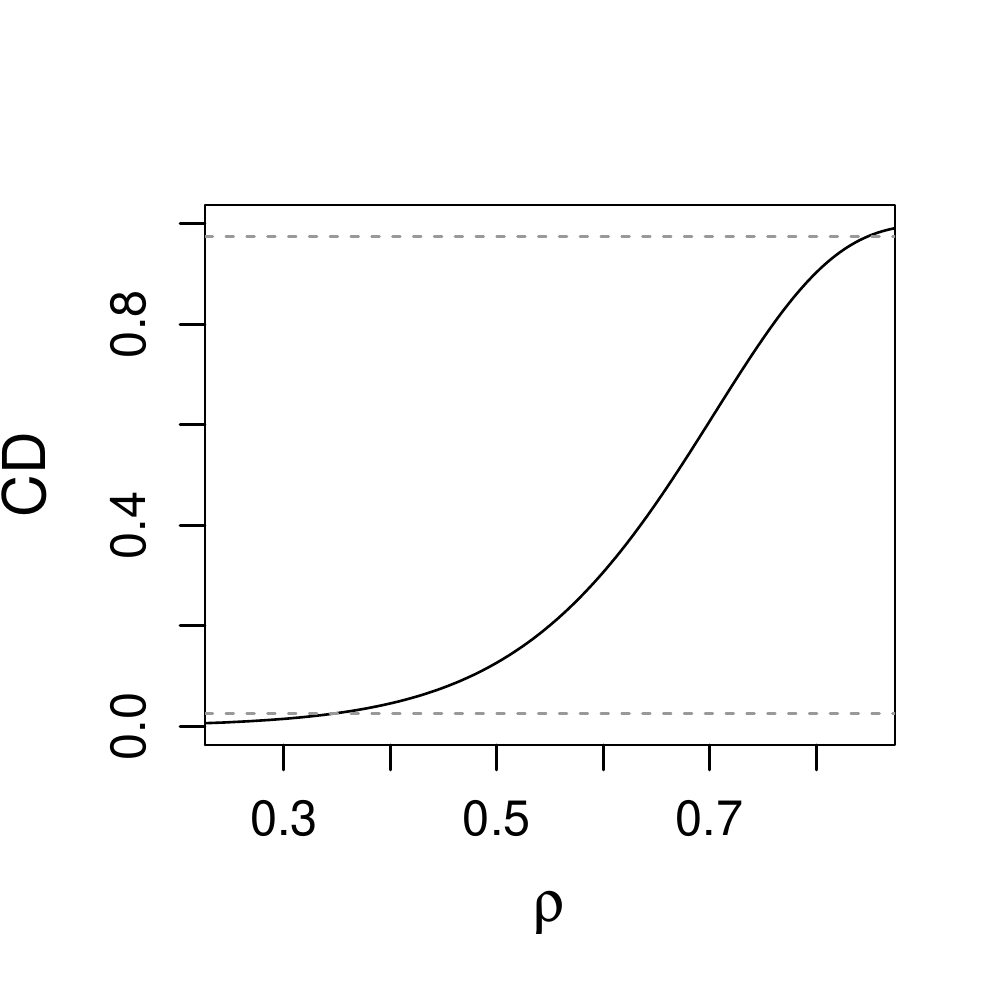}} \\
Bayes (uniform prior) &  (0.188,0.790) &  \raisebox{-.5\height}{\includegraphics[width=4.2cm]{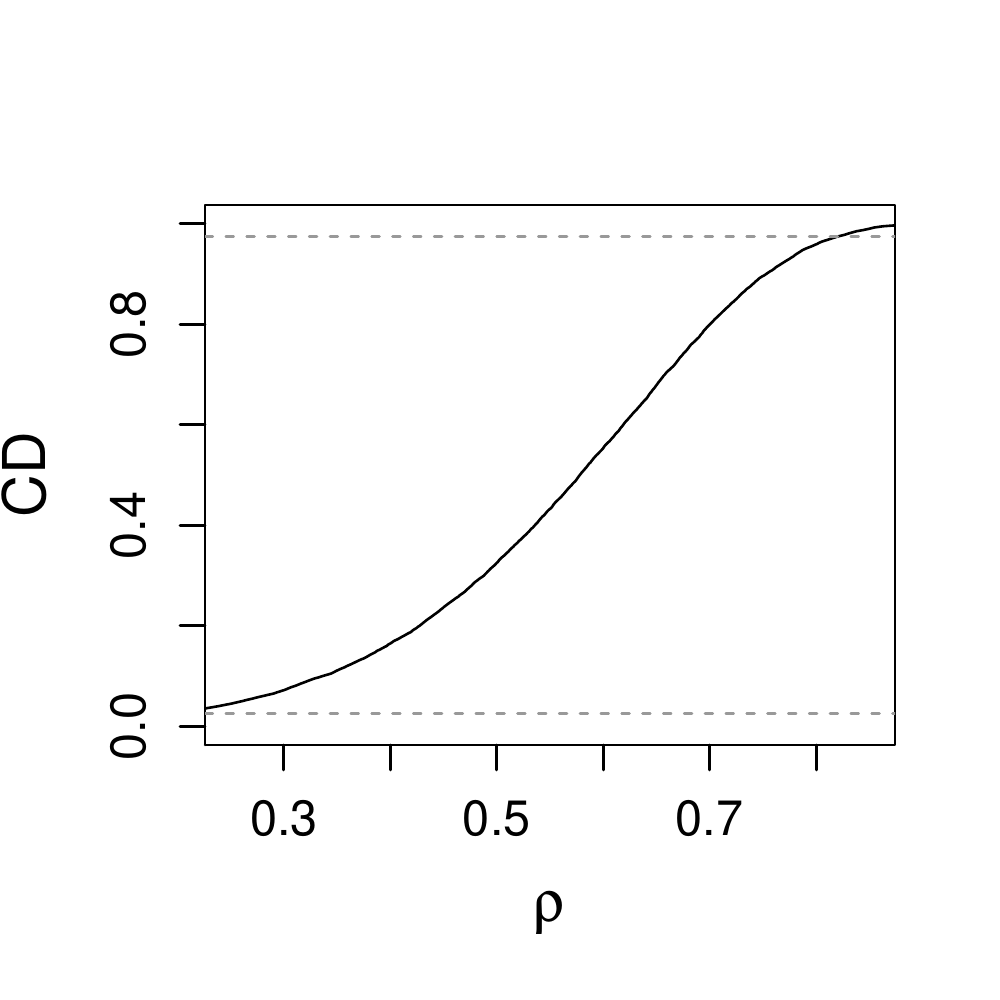}} \\
 \hline
Combination &  (0.505,0.760) & \raisebox{-.5\height}{\includegraphics[width=4.2cm]{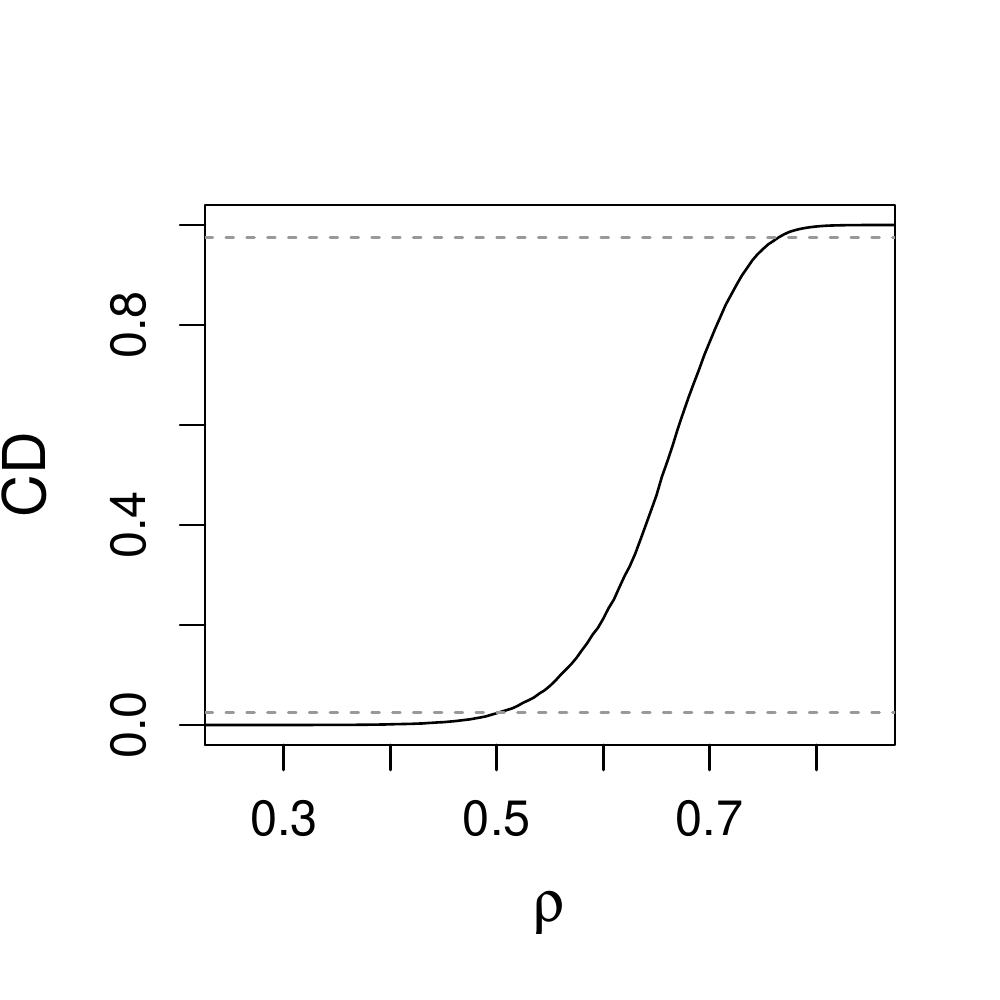}}\\
 \hline
\end{tabular}
\label{table1}
\end{table}

The results of the analysis are summarized in Table~\ref{table1}. As we can see from the table, four methods in different studies provide more or less similar results and the combined interval is much shorter than any of the four individual intervals.
In order to study the performance of the combination of \CDs in this situation, we present a simulation study with 200 replications.  Table~\ref{table2} shows the coverage and average length of 95\% CIs. We see that not only the combined approach maintains the desired coverage but also the length of CIs is roughly half of the lengths of CIs from individual studies. This result is as expected, since theoretically each study provides a $n^{-1/2}$-CIs and the sample size of combined data is $4n$ so we expect to obtain $(4n)^{-1/2}$-CI.

\begin{table}[h]
\center
\caption{Combination of four independent bivariate normal studies via \CDs}
\begin{tabular}{ccc}
 \hline
Methods & Coverage  & Mean length~(sd) of 95\% CIs \\
 \hline
Fisher's $Z$ method & 0.948 & 0.484~(0.140) \\
Bootstrap $\text{BC}_a$ & 0.936  & 0.464~(0.156)  \\
Profile likelihood & 0.918  & 0.436~(0.131)  \\
Bayes (uniform prior) & 0.964 &  0.522~(0.128) \\
 \hline
Combination & 0.954  & 0.226~(0.041)\\
 \hline
\end{tabular}
\label{table2} \\
\end{table}

\vspace{0.5cm}

Data \citep{diciccio1996}:

Baseline:
2.12,
4.35,
3.39,
2.51,
4.04,
5.10,
3.77,
3.35,
4.10,
3.35,
4.15,
3.56,
3.39,
1.88,
2.56,
2.96,
2.49,
3.03,
2.66,
3.00

One year:
2.47,
4.61,
5.26,
3.02,
6.36,
5.93,
3.93,
4.09,
4.88,
3.81,
4.74,
3.29,
5.55,
2.82,
4.23,
3.23,
2.56,
4.31,
4.37,
2.40

\vspace{2cm}
\textbf{\large Lists of abbreviations and symbols:}

\begin{itemize}
\item CD: confidence distribution

\item GFD: generalized fiducial distribution

\item CI: confidence interval

\item CV: confidence curve

\item $H_n(\cdot)$: confidence distribution

\item $H_n^{+}(\cdot), H_n^{-}(\cdot)$: upper and lower confidence distribution

\item $G$: data generating equation

\item $J$: Jacobian function
\end{itemize}

\newpage

\bibliographystyle{rss}
\bibliography{review}
\end{document}